\documentclass[]{aastex631}

\defcitealias{2023ApJS..266...28L}{L23}

\shortauthors{Li et al.}
\graphicspath{{./}{figures/}}
\usepackage{courier}
\usepackage{longtable}
\begin{document}

\title{Detection of Tidally Excited Oscillations in Kepler Heartbeat Stars}

\author[0000-0002-8564-8193]{Min-Yu Li}
\affiliation{Yunnan Observatories, Chinese Academy of Sciences, Kunming 650216, People's Republic of China}
\affiliation{Department of Astronomy, Key Laboratory of Astroparticle Physics of Yunnan Province, Yunnan University, Kunming 650091, People's Republic of China}
\affiliation{University of Chinese Academy of Sciences, No.1 Yanqihu East Road, Huairou District, Beijing 101408, People's Republic of China}
\affiliation{Center for Astronomical Mega-Science, Chinese Academy of Sciences, 20A Datun Road, Chaoyang District, Beĳing, 100012, People's Republic of China}

\author[0000-0002-5995-0794]{Sheng-Bang Qian}
\altaffiliation{E-mail: qsb@ynao.ac.cn}
\affiliation{Yunnan Observatories, Chinese Academy of Sciences, Kunming 650216, People's Republic of China}
\affiliation{Department of Astronomy, Key Laboratory of Astroparticle Physics of Yunnan Province, Yunnan University, Kunming 650091, People's Republic of China}
\affiliation{University of Chinese Academy of Sciences, No.1 Yanqihu East Road, Huairou District, Beijing 101408, People's Republic of China}
\affiliation{Center for Astronomical Mega-Science, Chinese Academy of Sciences, 20A Datun Road, Chaoyang District, Beĳing, 100012, People's Republic of China}

\author[0000-0002-0796-7009]{Li-Ying Zhu}
\altaffiliation{E-mail: zhuly@ynao.ac.cn}
\affiliation{Yunnan Observatories, Chinese Academy of Sciences, Kunming 650216, People's Republic of China}
\affiliation{University of Chinese Academy of Sciences, No.1 Yanqihu East Road, Huairou District, Beijing 101408, People's Republic of China}
\affiliation{Center for Astronomical Mega-Science, Chinese Academy of Sciences, 20A Datun Road, Chaoyang District, Beĳing, 100012, People's Republic of China}

\author[0000-0001-9346-9876]{Wen-Ping Liao}
\affiliation{Yunnan Observatories, Chinese Academy of Sciences, Kunming 650216, People's Republic of China}
\affiliation{University of Chinese Academy of Sciences, No.1 Yanqihu East Road, Huairou District, Beijing 101408, People's Republic of China}
\affiliation{Center for Astronomical Mega-Science, Chinese Academy of Sciences, 20A Datun Road, Chaoyang District, Beĳing, 100012, People's Republic of China}

\author{Er-Gang Zhao}
\affiliation{Yunnan Observatories, Chinese Academy of Sciences, Kunming 650216, People's Republic of China}
\affiliation{University of Chinese Academy of Sciences, No.1 Yanqihu East Road, Huairou District, Beijing 101408, People's Republic of China}
\affiliation{Center for Astronomical Mega-Science, Chinese Academy of Sciences, 20A Datun Road, Chaoyang District, Beĳing, 100012, People's Republic of China}

\author[0000-0002-5038-5952]{Xiang-Dong Shi}
\affiliation{Yunnan Observatories, Chinese Academy of Sciences, Kunming 650216, People's Republic of China}
\affiliation{University of Chinese Academy of Sciences, No.1 Yanqihu East Road, Huairou District, Beijing 101408, People's Republic of China}
\affiliation{Center for Astronomical Mega-Science, Chinese Academy of Sciences, 20A Datun Road, Chaoyang District, Beĳing, 100012, People's Republic of China}

\author[0000-0002-0285-6051]{Fu-Xing Li}
\affiliation{Yunnan Observatories, Chinese Academy of Sciences, Kunming 650216, People's Republic of China}
\affiliation{University of Chinese Academy of Sciences, No.1 Yanqihu East Road, Huairou District, Beijing 101408, People's Republic of China}
\affiliation{Center for Astronomical Mega-Science, Chinese Academy of Sciences, 20A Datun Road, Chaoyang District, Beĳing, 100012, People's Republic of China}

\author[0000-0003-0516-404X]{Qi-Bin Sun}
\affiliation{Yunnan Observatories, Chinese Academy of Sciences, Kunming 650216, People's Republic of China}
\affiliation{University of Chinese Academy of Sciences, No.1 Yanqihu East Road, Huairou District, Beijing 101408, People's Republic of China}
\affiliation{Center for Astronomical Mega-Science, Chinese Academy of Sciences, 20A Datun Road, Chaoyang District, Beĳing, 100012, People's Republic of China}

\author[0009-0004-0289-2732]{Ping Li}
\affiliation{Yunnan Observatories, Chinese Academy of Sciences, Kunming 650216, People's Republic of China}
\affiliation{University of Chinese Academy of Sciences, No.1 Yanqihu East Road, Huairou District, Beijing 101408, People's Republic of China}
\affiliation{Center for Astronomical Mega-Science, Chinese Academy of Sciences, 20A Datun Road, Chaoyang District, Beĳing, 100012, People's Republic of China}



\begin{abstract}

Heartbeat stars (HBSs) with tidally excited oscillations (TEOs) are ideal laboratories for studying the effect of equilibrium and dynamical tides. However, studies of TEOs in Kepler HBSs are rare due to the need for better modeling of the equilibrium tide in light curves. We revisit the HBSs in our previous work and study the TEOs in these HBSs based on the derived orbital parameters that could express the equilibrium tide. We also compile a set of analytic procedures to examine the harmonic and anharmonic TEOs in their Fourier spectra. The TEOs of 21 HBSs have been newly analyzed and presented. Twelve of these HBSs show prominent TEOs (signal-to-noise ratio of the harmonics $S/N \ge 10$). The relation between the orbital eccentricities and the harmonic number of the TEOs shows a positive correlation. The relation between the orbital periods and the harmonic number also shows a positive correlation. Furthermore, the distribution of HBSs with TEOs in the Hertzsprung-Russell (H-R) diagram shows that TEOs are more visible in hot stars with surface effective temperatures $T$ $\gtrsim$ 6500 K. These samples may also be valuable targets for future studies of the effect of tidal action in eccentric orbits.

\end{abstract}



\section{Introduction} \label{sec:intro}
Heartbeat stars (HBSs) are named for the shape of their light curve, which resembles an echocardiogram \citep{2012ApJ...753...86T}. HBSs with eccentric orbits are distorted by the time-variable tidal potential of the companion star, and the response of the star is usually divided into two components: the equilibrium tide and the dynamical tide. The equilibrium tide contributes to the ``heartbeat" signature near periastron, while the dynamical tide consists of tidally excited oscillations (TEOs; \citet{1975A&A....41..329Z,1995ApJ...449..294K}) that are visible in all orbital phases \citep{2017MNRAS.472.1538F}. Stellar oscillations are an important source of information about the internal properties of HBSs, so HBSs with TEOs are ideal laboratories for studying the effect of equilibrium and dynamical tides \citep{2020ApJ...888...95G}.

Theoretical work on TEOs is extensive. \citet{2017MNRAS.472.1538F} made detailed theoretical predictions for the photometric amplitudes and phases of TEOs, and the tidal dissipation they produce. \citet{2021FrASS...8...67G} discussed the theoretical approaches used to model TEOs, and the effect of mass transfer, the extreme result of tides, on stellar oscillations. \citet{2023A&A...671A..22K} investigated the details of TEOs in massive and intermediate-mass stars based on theoretical modeling combined with machine learning (ML) techniques. However, due to the very small photometric variations, such oscillations are difficult to detect from ground-based observations in the pre-Kepler era \citep{2017MNRAS.472.1538F}.

The NASA Kepler space telescope \citep{2010Sci...327..977B} provides high-precision photometric data with a long time baseline that are ideal for studying HBSs. The prototype HBS KIC 8112039 (KOI-54) was first reported by \citet{2011ApJS..197....4W} based on its Kepler light curve and spectroscopic observations. The two dominant TEOs have the 90th and 91st harmonics of the orbital frequency. KOI-54 was then studied in a series of papers, including \citet{2012MNRAS.421..983B}, \citet{2012MNRAS.420.3126F}, \citet{2014MNRAS.440.3036O}, and \citet{2022MNRAS.517..437G}.

In addition, other systems exhibit TEOs. \citet{2012ApJ...753...86T} reported 17 HBSs with evidence of TEOs. \citet{2013MNRAS.434..925H} reported KIC 4544587 with $\delta$ Scuti ($\delta$ Sct) pulsations and TEOs. \citet{2016MNRAS.463.1199H} reported KIC 3749404 with rapid apsidal advance and TEOs. \citet{2017MNRAS.472L..25F, 2018MNRAS.473.5165H} reported the resonance locking in KIC 8164262. \citet{2017ApJ...834...59G, 2020ApJ...896..161G} examined KIC 3230227 with evidence of nonlinear mode coupling. \citet{2019ApJ...885...46G} studied the TEOs, $\delta$ Sct p modes and the self-driven $\gamma$ Doradus-type ($\gamma$ Dor) g modes in KIC 4142768. \citet{2020ApJ...888...95G} examined the pulsation phases of TEOs in eight HBSs. \citet{2020ApJ...903..122C} characterized the TEOs in three Kepler HBSs, and found that resonance locking is likely occurring in KIC 11494130, but not in KIC 6117415 or KIC 5790807. \citet{2019MNRAS.489.4705J, 2021MNRAS.506.4083J, 2022A&A...659A..47K} reported the TEOs in MACHO 80.7443.1718. \citet{2021A&A...647A..12K} discovered 20 massive HBSs from TESS, seven of which have TEOs. \citet{2022ApJ...928..135W, 2022ApJS..259...16W} reported 991 OGLE HBSs and found TEOs in 52 systems. \citet{2023AJ....166...42W} reported the linear and nonlinear TEOs and $\delta$ Sct pulsations in HBS FX UMa.

When studying TEOs in HBSs, the contribution of the equilibrium tide must first be subtracted from the light curves. The remaining dynamical tide, which contains features of TEOs, can then be used to perform Fourier spectral analysis \citep{2020ApJ...888...95G}. \citet{2016AJ....151...68K} have reported a catalog of 173 Kepler HBSs, of which 24 have been flagged as TEOs. So far, only ten or more of these HBSs with TEOs have been studied in detail, due to the need for better modeling of the equilibrium tide in the light curves. In the first paper in this series (\citet{2023ApJS..266...28L}, hereafter \citetalias{2023ApJS..266...28L}), we have modeled 153 Kepler HBSs in the \citet{2016AJ....151...68K} catalog using a corrected version of the \citet{1995ApJ...449..294K} model (K95$^+$ model), which represents the equilibrium tide of HBSs, and flagged a number of them as TEOs by visual inspection. However, the signature of TEOs is that they occur at exact integer multiples of the orbital frequency, which distinguishes them from self-excited oscillations and stochastically excited pulsations \citep{2017MNRAS.472.1538F}. Therefore, based on the orbital parameters derived in \citetalias{2023ApJS..266...28L}, we revisit the HBSs to confirm these TEOs and search for other TEOs that could not be found by visual inspection. We re-examine the harmonic TEOs and anharmonic TEOs of these systems, where the multimode coupling includes not only second-order coupling but also third-order coupling \citep{2014MNRAS.440.3036O, 2021FrASS...8...67G}. After re-examining 146 HBSs (excluding the seven systems that should be excluded from the HBS candidates in Section 5.2 of \citetalias{2023ApJS..266...28L}), we identify 21 HBSs with TEOs, excluding eight systems studied in previous works. 

This paper presents the analytic procedures and results for these 21 systems. Section \ref{sec:data_analysis} describes the photometric data and other astrometric data used in this work and the analytic procedure for the TEOs. Section \ref{sec:rst} presents the results. Section \ref{sec:discussion} discusses the reliability of the results of analysis, the Hertzsprung-Russell (H-R) diagram of these HBSs and the parameter statistics. Section \ref{sec:conclusions} summarizes and concludes our work.

\section{Data and Analyses} \label{sec:data_analysis}
\subsection{Data Reduction}
The Kepler Q0 to Q17 mission provided essentially uninterrupted and ultrahigh precision photometric data within a 105 deg$^2$ field of view in the constellations of Cygnus and Lyra from 2009 May to 2013 May. This work uses the Kepler photometric data obtained in \citetalias{2023ApJS..266...28L}. It consists of two parts, one provided in the \citet{2016AJ....151...68K} catalog\footnote{\url{http://keplerEBs.villanova.edu}} and the other downloaded from the Kepler public data using the {\tt\string lightkurve} package \citep{2018ascl.soft12013L} (details in \citetalias{2023ApJS..266...28L}).

The Large Sky Area Multi-Object Fiber Spectroscopic Telescope (LAMOST, also called the Guo Shou Jing Telescope) is a special Schmidt reflecting telescope: it has a large field of view (5$^{\circ}$) and can obtain 4000 spectra in a single exposure \citep{1996ApOpt..35.5155W, 2012RAA....12.1197C}. It has obtained a large number of stellar spectral data \citep{2012RAA....12..723Z, 2015RAA....15.1095L}. 

The Gaia satellite, a cornerstone mission of the European Space Agency launched in 2013 December, has provided very high-precision astrometric data, including parallax data, visual magnitude, and the stellar atmospheric parameters for nearly 2 billion stars \citep{2016A&A...595A...1G, 2018A&A...616A...1G, 2021A&A...649A...1G}.

\subsection{Detection of TEOs}
Since Fourier spectral analysis is computationally intensive for such a large number of HBSs, we use the FNPEAKS \footnote{\url{http://helas.astro.uni.wroc.pl/deliverables.php?active=fnpeaks}} code to perform the Fourier transform of the photometric data. It uses the modification proposed by \citet{1985MNRAS.213..773K} and substantially reduces the computational time for a Fourier transform. Uncertainties of frequencies and amplitudes are estimated following \citet{2008A&A...481..571K}. To calculate the signal-to-noise ratio ($S/N$) of the frequencies, we derive the mean noise level, $N$, as the mean amplitude in the frequency range $\pm$0.5 day$^{-1}$, and include only the frequencies with $S/N \ge 4.0$ in the analysis. 

The frequency $f$ is considered a harmonic TEO candidate if it satisfies at least one of the following equations:
\begin{equation}\label{equation:a}
	|n-f/f_{\rm orb}| < 0.01,
\end{equation}
\begin{equation}\label{equation:b}
	|n-f/f_{\rm orb}| < 3\sigma_{f/f_{\rm orb}}, 
\end{equation}
where $n$ is the harmonic number, $f_{\rm orb}=1/P$ is the orbital frequency, $\sigma_{f/f_{\rm orb}}=\sqrt{P^2\sigma_f^2+f^2\sigma_P^2}$ is the error of $f/f_{\rm orb}$ according to \citet{2021A&A...647A..12K} and \citet{2022ApJ...928..135W}, $\sigma_f$ and $\sigma_P$ stand for the errors of $f$ and $P$, respectively. $P$ and $\sigma_P$ are from \citetalias{2023ApJS..266...28L} using the Markov Chain Monte Carlo method. 

We suggest that these two equations are appropriate for different scenarios. Eq. (\ref{equation:a}) is appropriate for most frequencies from high-precision Kepler photometric data, while Eq. (\ref{equation:b}) is not easily satisfied since a higher $S/N$ and a large number of data points lead to a lower $\sigma_f$ according to the approach to uncertainty estimation of \citet{2008A&A...481..571K} (their Eq. (4)). However, as the harmonic number $n$ increases, the frequency error is amplified, making Eq. (\ref{equation:a}) unsatisfiable, and Eq. (\ref{equation:b}) should be considered. The harmonic $n$ = 1118 of KIC 5006817 in Table \ref{tab:TEOs} is an example. In addition, for the photometric data with a small number of data points or lower precision, Eq. (\ref{equation:b}) is also appropriate \citep{2021A&A...647A..12K, 2022ApJ...928..135W}.

In addition, we also examine anharmonic TEOs where the multimode coupling includes both second-order and third-order coupling, since the high-precision Kepler photometric data are sufficient for more in-depth analysis. In these cases, the frequency $f$ in Eqs. (\ref{equation:a}) and (\ref{equation:b}) should be derived as $f$=$f_1$+$f_2$ or $f$=$f_1$+$f_2$+$f_3$, where $f_1$, $f_2$, and $f_3$ are anharmonic frequencies (see Figure 5 in \citet{2021FrASS...8...67G}). The uncertainties are calculated analogously using the error propagation equation.

\begin{figure}
	\centering
	\includegraphics[width=1.0\textwidth]{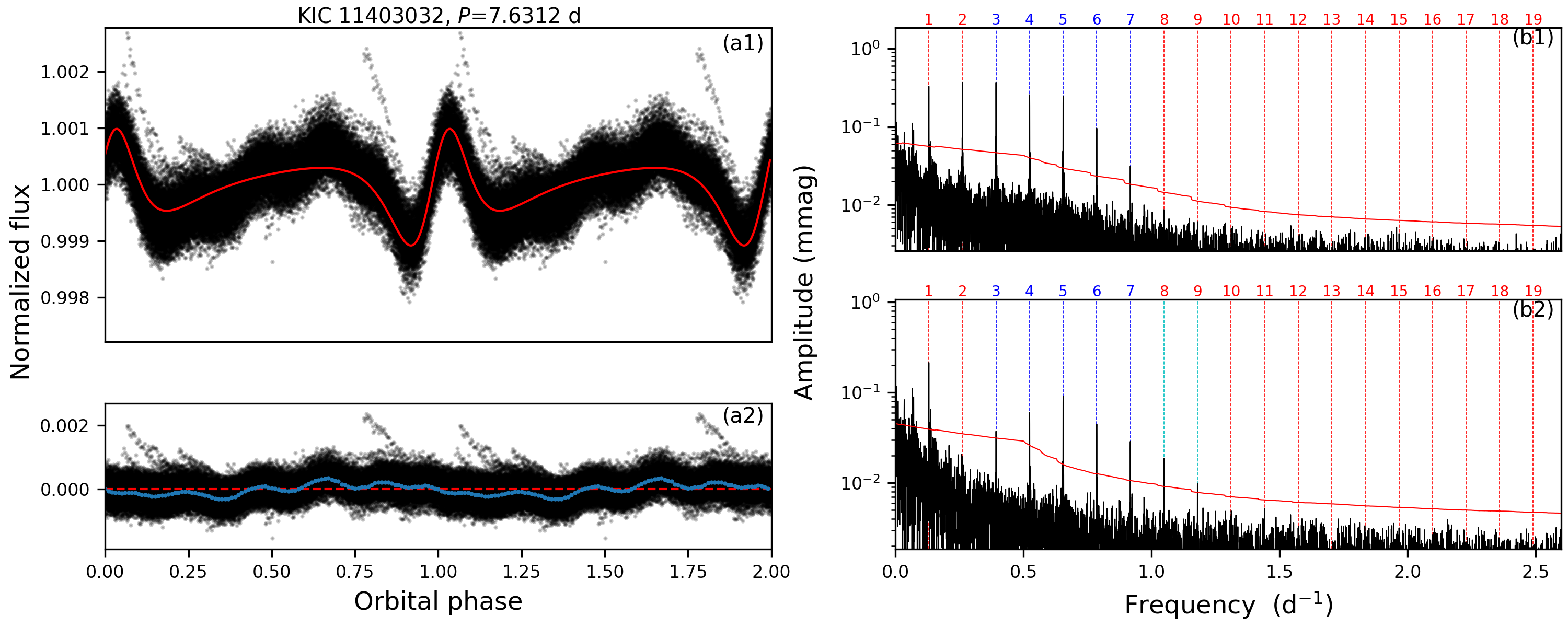}
	\caption{The analytic procedure for KIC 11403032. (a1) The K95$^+$ model (solid red line) fitted to the phase-folded light curve (black dots). Note that phases (1) and (2) are an exact copy of phases (0) and (1) to show the TEOs more clearly. (a2) The residuals of the fit in panel (a1). The blue dots are medians in 0.01 phase bins. (b1) The frequency spectrum of the light curve shown in panel (a1). (b2) The frequency spectrum of the residuals shown in panel (a2). Panels (b1) and (b2): The solid red line shows the amplitudes at $S/N$ = 4.0 as a function of frequency; the red, cyan, and blue vertical dashed lines represent the orbital harmonics $n$; the blue ones indicate that they are TEOs, but the cyan ones are not (details in text).
	\label{fig:11403032-ft}}
\end{figure}

The analytic procedure for KIC 11403032 is shown in Figure \ref{fig:11403032-ft} as an example. The phase-folded light curve and the results of the K95$^+$ model fit, derived in \citetalias{2023ApJS..266...28L}, are shown in panel (a1). The fit residuals are shown in panel (a2), and the blue dots are added as medians in 0.01 phase bins. We apply Fourier transforms not only to the residuals (panel (b2)), but also to the original light curve (panel (b1)). Note that any processing of the original data before Fourier transformation may produce fictitious harmonics in the spectrum, because it is not certain that the model is a perfect fit. Therefore, when examining harmonic TEOs, we also apply Fourier transform to the original light curve, and combine all Fourier spectra for analysis. It will be considered a harmonic TEO if a harmonic $n$ occurs in all spectra. We also visually inspect all fits to reduce the possibility that the model subtraction injects orbital harmonic frequencies into the residuals. In this case, harmonics $n$ = 8 and 9 occur in panel (b2), but are not significant in panel (b1). As a result, we consider the harmonic $n$ to be a TEO if it occurs in both panels (b1) and (b2), and we represent it as a blue dashed line. Otherwise, if the harmonic $n$ occurs in only one panel, it would be represented as a cyan dashed line in that panel. KIC 11403032 is a typical example, and although such a case is unusual in our samples, we suggest an analysis of the Fourier spectrum of both the original data and the residuals. In addition, we suggest that the $n$ = 1 and 2 harmonics are artifacts of imperfect modeling or detrending and should be ignored in all cases. Finally, the TEO candidates are the $n$ = 5, 4, 6, 3, and 7 harmonics.

\begin{figure}
	\centering
	\includegraphics[width=1.0\textwidth]{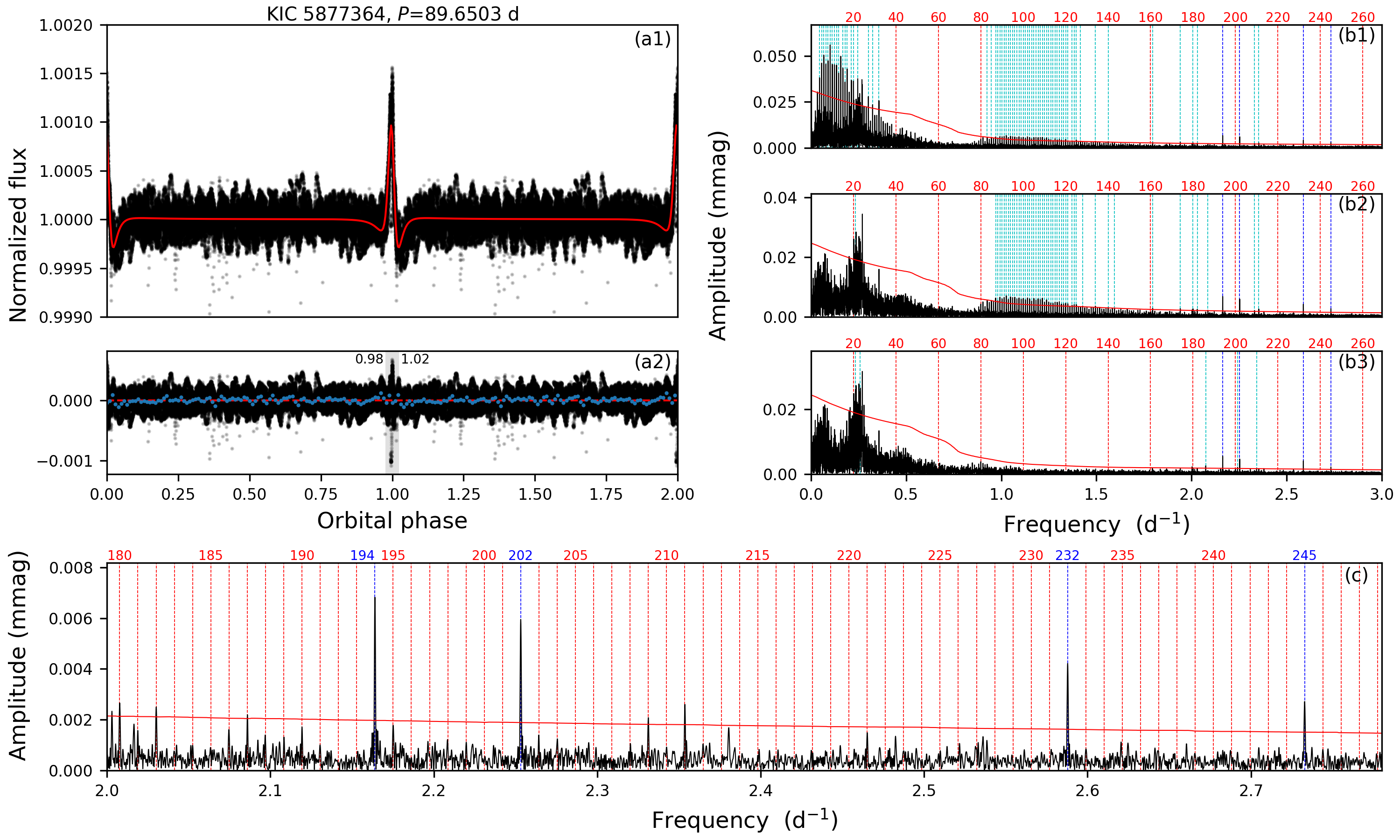}
	\caption{The analytic procedure for KIC 5877364. Panels (a1), (a2), (b1), and (b2) are the same as in Fig. \ref{fig:11403032-ft}, but for KIC 5877364. (b3) The frequency spectrum of the residuals is shown in panel (a2), with the data at the phases marked by the gray stripe removed. (c) Zoom-in on panel (b3).
		\label{fig:5877364-ft}}
\end{figure}

Furthermore, if the remnants exist near the periastron passage in the residuals, we remove them and apply the Fourier transform again. Take KIC 5877364 as an example, shown in Figure \ref{fig:5877364-ft}. As seen in panel (a2), remnants of the heartbeat and/or eclipsing remain in the residuals near the periastron passage. These remnants can sometimes mimic TEOs because a narrow heartbeat produces high orbital harmonics \citep{2021A&A...647A..12K}. Therefore, we remove the data at the phases marked by the gray stripe (the phase range is marked next to it in panel (a2)) and reapply a Fourier transform to obtain a new spectrum, which is shown in panel (b3). The harmonic $n$ is considered a TEO and is marked as a blue dashed line if it appears in all three panels. Otherwise, it is marked as a cyan dashed line. In this example, the harmonics in the range of 0.9 day$^{-1}$ $\sim$ 1.6 day$^{-1}$ (cyan lines in panels (b1) and (b2)) are fictitious harmonics and should be ignored. 

We examine all 146 HBSs using the analytic procedure described above. Although we cannot completely rule out fictitious frequencies, we suggest such a practical approach.

\section{Results} \label{sec:rst}
The examination in Section \ref{sec:data_analysis} shows that 21 HBSs exhibit TEOs. We present the most prominent harmonic and anharmonic TEOs for each HBS in Table \ref{tab:TEOs}. The entire table is available in machine-readable form. In addition, a harmonic TEO is classified as a prominent harmonic TEO if its $S/N \ge 10$; otherwise, it is classified as a non-prominent TEO. We mark 12 prominent TEO HBSs with a superscript ``$a$" next to the KIC ID. The analytic plots for systems other than KIC 11403032 and KIC 5877364 are shown in Appendix \ref{sec:apdx}.

\startlongtable
\begin{deluxetable*}{lrrrrrrrrc}
	\tabletypesize{\scriptsize}
	\label{tab:TEOs}
	\tablenum{1}
	\tablecaption{Harmonic and anharmonic TEOs of 21 HBSs (Partially).}
	\tablehead{
		\colhead{KIC}& \colhead{$e$} & \colhead{$n$} & \colhead{$\Delta n$} &  \colhead{$3\sigma-|\Delta n|$} & \colhead{Frequency} & \colhead{Amplitude}  & \colhead{$f/f_{\rm orb}$} & \colhead{$S/N$} & \colhead{Remark}\\
		{} & {} & {} & {} &{} & \colhead{(day$^{-1}$)} & \colhead{(mmag)} &{} 
	} 
	\startdata
3547874 $^a$ & 0.648 & 37 & 0.000 & 0.000218 & 1.8789330(72) & 0.03602(69) & 37.000 & 32.44 \\ 
\hline
3862171 & 0.594 & 7 & 0.007 & -0.003811 & 0.333850(49) & 0.0173(22) & 7.007 & 4.81 \\ 
\hline
4377638 $^a$ & 0.220 & 7 & 0.000 & -0.000365 & 2.4808723(62) & 0.07354(96) & 7.000 & 47.61 \\ 
\hline
4949187 $^a$ & 0.545 & 11 & 0.000 & 0.000296 & 0.918301(21) & 0.0593(33) & 11.000 & 10.98 \\ 
\hline
5006817 & 0.713 & 1118 & -0.013 & 0.007475 & 11.787470(61) & 0.0211(18) & 1117.987 & 7.30 \\ 
\hline
5090937 $^a$ & 0.249 & 6 & 0.000 & 0.000149 & 0.6818264(60) & 0.1247(20) & 6.000 & 38.88 \\ 
\hline
5129777 $^a$ & 0.708 & 102 & 0.003 & -0.000340 & 3.899310(15) & 0.01890(75) & 102.003 & 15.62 \\ 
\cline{3-10}
{} & {} & 102 & 0.000 & 0.003329 & 2.6048160(65) & 0.1379(24) & 68.140 & 35.78 & $\mathit{f_{1}}$ \\ 
\cline{6-10}
{} & {} & {} & {} & {} & 1.294392(33) & 0.00963(85) & 33.860 & 7.03 & $\mathit{f_{2}}$ \\ 
\hline
5877364 $^a$ & 0.719 & 194 & 0.001 & 0.003633 & 2.163979(18) & 0.00567(28) & 194.001 & 12.70 \\ 
\hline
5960989 & 0.811 & 77 & -0.004 & 0.002522 & 1.518400(45) & 0.0102(12) & 76.996 & 5.10 \\ 
\hline
6290740 & 0.215 & 5 & 0.002 & -0.000564 & 0.330134(31) & 0.0318(26) & 5.002 & 7.46 \\ 
\hline
7918217 & 0.906 & 37 & 0.003 & 0.004916 & 0.578764(40) & 0.00772(82) & 37.003 & 5.82 \\ 
\hline
8264510 $^a$ & 0.341 & 7 & 0.000 & 0.000068 & 1.2309394(65) & 0.1965(34) & 7.000 & 35.49 \\ 
\hline
8456774 $^a$ & 0.275 & 8 & 0.000 & -0.000003 & 2.7717813(44) & 0.3558(41) & 8.000 & 53.17 \\ 
\hline
8459354 $^a$ & 0.941 & 99 & -0.002 & 0.002020 & 1.848514(27) & 0.1492(57) & 98.998 & 16.07 \\ 
\cline{3-10}
{} & {} & 1249 & -0.002 & 0.013779 & 23.321668(95) & 0.0495(67) & 1248.998 & 4.56 \\ 
\cline{3-10}
{} & {} & 1249 & -0.005 & 0.009409 & 14.278236(15) & 0.744(16) & 764.675 & 28.47 & $\mathit{f_{1}}$ \\ 
\cline{6-10}
{} & {} & {} & {} & {} & 9.043368(88) & 0.0358(45) & 484.320 & 4.94 & $\mathit{f_{2}}$ \\ 
\hline
8838070 & 0.802 & 15 & 0.000 & 0.005878 & 0.345806(48) & 0.0187(24) & 15.000 & 4.90 \\ 
\hline
9535080 & 0.816 & 10 & 0.009 & -0.000984 & 0.201603(52) & 0.0087(12) & 10.009 & 4.45 \\ 
\hline
9717958 & 0.734 & 23 & 0.004 & 0.005847 & 0.338851(41) & 0.0205(22) & 23.004 & 5.72 \\ 
\hline
9972385 & 0.877 & 232 & 0.004 & 0.001480 & 3.971234(30) & 0.00679(55) & 232.004 & 7.61 \\ 
\hline
11122789 $^a$ & 0.327 & 6 & -0.001 & -0.000973 & 1.8525464(42) & 0.04136(46) & 5.999 & 55.45 \\ 
\hline
11403032 $^a$ & 0.294 & 5 & 0.000 & -0.000044 & 0.655168(10) & 0.0911(25) & 5.000 & 22.63 \\ 
\hline
11572363 $^a$ & 0.705 & 50 & 0.000 & 0.000009 & 2.6276087(82) & 0.04250(93) & 50.000 & 28.37 \\ 
\hline
	\enddata
	\tablecomments{Column 2 is the KIC ID; column 2 is the eccentricity derived in \citetalias{2023ApJS..266...28L}; $n$ is the harmonic number of the TEO; $\Delta n=f/f_{\rm orb}-n$, where $f$ is the detected frequency (column 6); a positive value in column 5 indicates that it satisfies Eq. (\ref{equation:b}); column 7 is the amplitude; $S/N$ is the signal-to-noise ratio; the Remark column gives the frequency in the anharmonic TEO, also labeled in panel (c) of the analytic procedure figures A8 and A14 (see Appendix \ref{sec:apdx}).\\
	$^a$ Indicates that the system has prominent harmonic TEOs ($S/N \ge 10$).\\
	(Only the most prominent harmonic and anharmonic TEOs are listed for each sample. The complete information is available in machine-readable form.)}
\end{deluxetable*}

\section{Discussions} \label{sec:discussion}
\subsection{Reliability of the Analytic Results}
\begin{figure}
	\centering
	\includegraphics[width=1.0\textwidth]{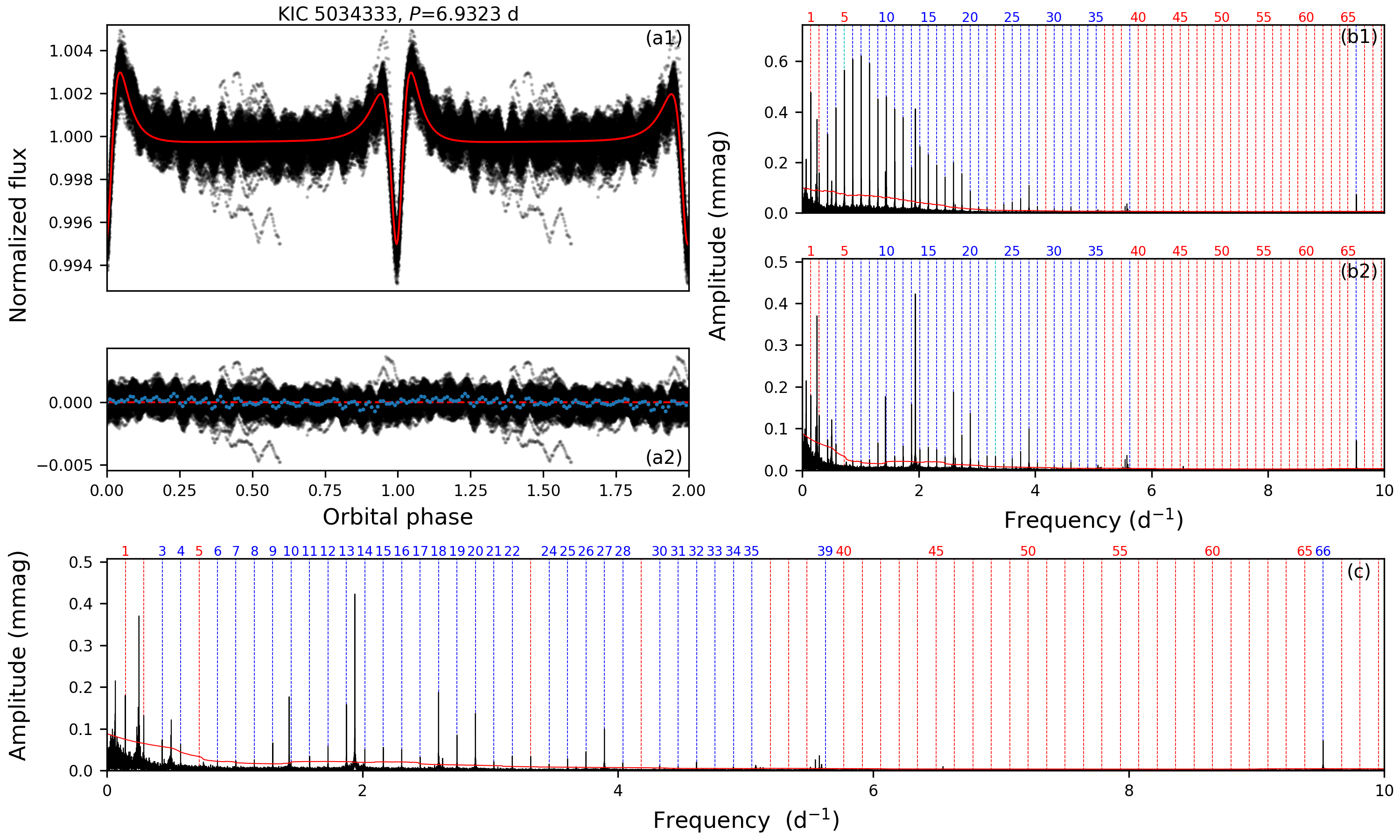}
	\caption{The analytic procedure for KIC 5034333. Panels (a1), (a2), (b1), and (b2) are the same as in Fig. \ref{fig:11403032-ft}, but for KIC 5034333. (c) Zoom-in on panel (b2).
		\label{fig:5034333-ft}}
\end{figure}

The FNPEAKS code is open-source software with excellent performance. It produces results consistent with other approaches (e.g., Period04, SigSpec, AOV software, and Lomb-Scargle periodogram) \citep{2021MNRAS.507..781N, 2021ApJS..255....4P, 2022ApJS..260...46I, 2019ApJ...881L..41P, 2022ApJS..259...16W}. To confirm the correctness of the analytic procedure, we perform a TEO analysis on KIC 5034333 as an example and compare the results with previous works. \citet{2020ApJ...888...95G} (G20) studied the eight prominent TEOs of KIC 5034333. Figure \ref{fig:5034333-ft} shows our analytic procedure for KIC 5034333. Table \ref{tab:compare} shows the eight most prominent TEOs derived from G20 and this work (other TEOs with lower amplitude are ignored). Our results are generally consistent with those of G20. We suggest that the minor differences in amplitude are due to different modeling of the equilibrium tide. As a result, we suggest that the results of our analytic procedure are reliable.

\startlongtable
\begin{deluxetable*}{rccrccccc}
	\tabletypesize{\scriptsize}
	\tablenum{2}
	\tablecaption{Comparison of the eight TEOs in KIC 5034333 with G20. \label{tab:compare}}
	\tablehead{
		\colhead{}& \colhead{} & \colhead{G20} & \colhead{} & \colhead{} & \colhead{} & \colhead{This work} & \colhead{} & \colhead{}\\
		\cline{2-3} \cline{5-9} 
		\colhead{$n$} & \colhead{Frequency(day$^{-1})$} &  \colhead{Amplitude(mmag)} & \colhead{} & \colhead{Frequency(day$^{-1})$} & \colhead{Amplitude(mmag)} & \colhead{$\Delta n$} &  \colhead{$3\sigma-|\Delta n|$} & \colhead{$S/N$}
	} 
	\startdata 
18	&	2.596525(2)	&	0.17600(15)	&{}&	2.5965304(44) & 0.1874(22) & -0.0001 & -0.0000242 & 53.13 \\ 
13	&	1.875299(4)	&	0.15005(28)	&{}&	1.8752607(78) & 0.1579(33) & -0.0002 & -0.0000004 & 29.93 \\
20	&	2.885043(3)	&	0.14651(21)	&{}&	2.8850452(53) & 0.1367(19) & 0.00005 & 0.0000609 & 43.81 \\
27	&	3.894829(4)	&	0.08778(21)	&{}&	3.8948297(41) & 0.0995(11) & 0.00006 & 0.0000223 & 56.49 \\ 
19	&	2.740803(5)	&	0.08017(23)	&{}&	2.7408048(91) & 0.0846(21) & 0.00004 & 0.0001535 & 25.50 \\
66	&	9.521462(2)	&	0.07225(3)	&{}&	9.5214617(34) & 0.07120(64) & 0.005 & -0.0054005 & 69.11 \\ 
4	&	0.576985(20)	&	0.06127(73)	&{}&	0.576961(41) & 0.0627(69) & -0.0003 & 0.0005140 & 5.65 \\
12	&	1.731012(9)	&	0.06019(5)	&{}&	1.731054(21) & 0.0581(33) & 0.0002 & 0.0002872 & 10.81 \\
		\enddata
	\tablecomments{The harmonic numbers $n$ are ordered as Table 4 in G20; columns 2$-$3 are derived from G20;  columns 4$-$8 are derived in this work.}
\end{deluxetable*}

\subsection{The H-R Diagram of These HBSs}
\label{subsec:HR}
To illustrate the H-R diagram of the 146 HBSs, the parallax, visual magnitude, and surface effective temperature are required. All of these 146 HBSs are observed by the Gaia Survey, from which we obtain the parallax, visual magnitude, and Gaia ID for each system. With the Gaia ID, we cross-match with the LAMOST catalogs and obtain the effective temperature for most systems. For other systems that are not in the LAMOST catalogs, we obtain the effective temperature from the Gaia Survey and \citet{2014ApJS..211....2H}. 

The luminosities of these samples are calculated using the following equations  \citep{2023ApJS..265...33S}:
\begin{equation}\label{equation:c}
	{\rm log}(L/L_{\odot})=0.4\cdot(4.74-M_V-BC)
\end{equation}
\begin{equation}\label{equation:d}
	M_V=m_V-5\cdot {\rm log}(1000/\pi)+5-A_V,
\end{equation}
where the bolometric correction BC is estimated following \citet{2023ApJS..265...33S} and \citet{2013ApJS..208....9P}; the visual magnitude $m_V$ and parallax $\pi$ are from the Gaia Survey; the interstellar extinction $A_V$ is 0.22 mag for the Kepler field \citep{2017MNRAS.470L..97G}.

\begin{figure}
	\gridline{
		\fig{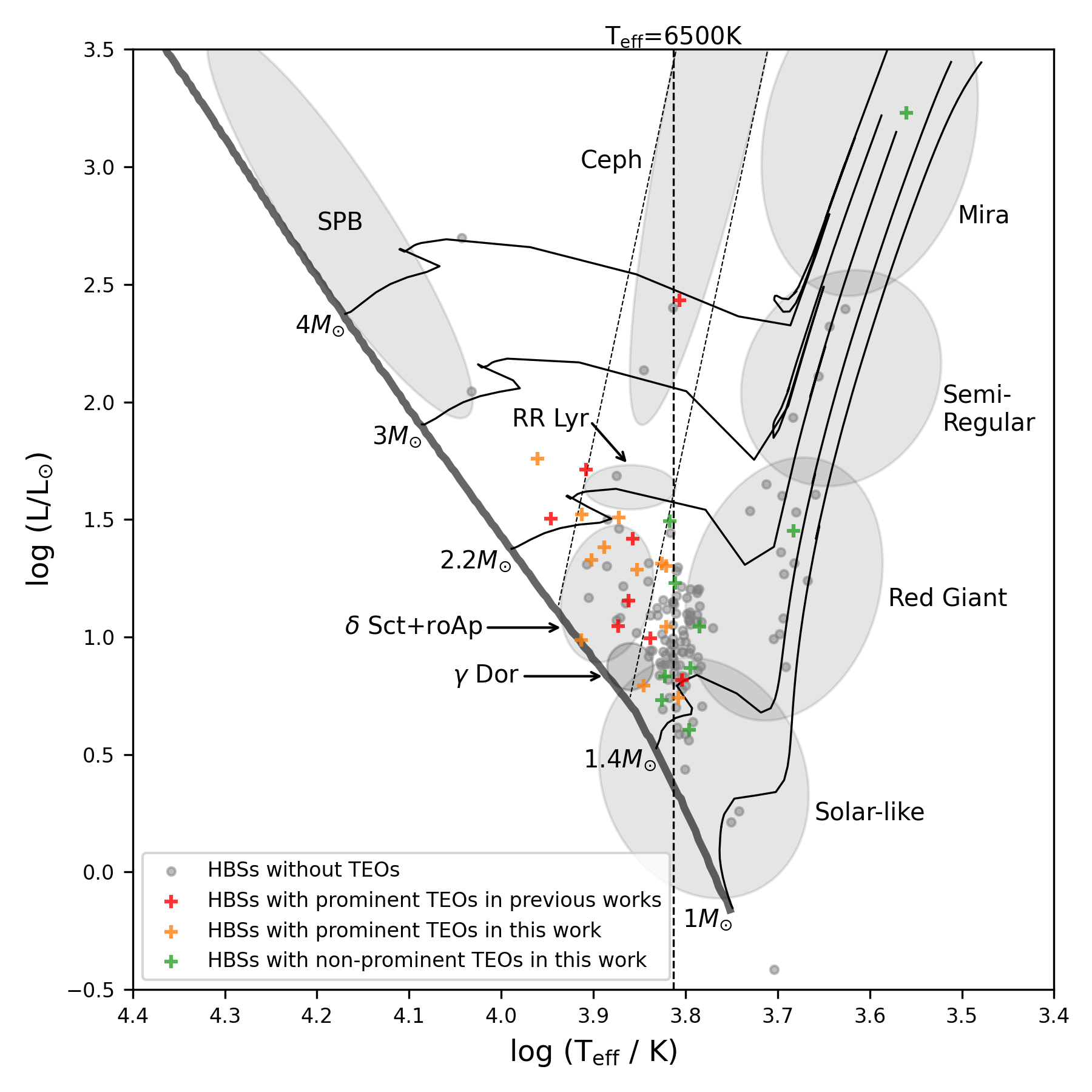}{0.75\textwidth}{}
	}
	\caption{The H-R diagram of the 146 HBSs. The red pluses represent eight HBSs with TEOs mentioned in Section \ref{subsec:HR}. The orange pluses represent 12 HBSs with prominent TEOs, and the green pluses represent nine HBSs without prominent TEOs in this work. The gray circles represent the other 117 HBSs without TEOs. The bold black solid curve represents the ZAMS. The black lines are the theoretical evolutionary tracks for the different masses with Z = 0.02. The gray dashed curves show the classical instability strip for radial pulsations. The classes of pulsating variable stars, including SPB, Ceph, RR Lyr, $\delta$ Sct+roAp, $\gamma$ Dor, Mira, semi-regular, red giant, and solar-like, are labeled next to the corresponding gray regions. Figure design from \citet{2021RvMP...93a5001A, 2019ApJS..243...10P, 2013PhDT.........6P}.
		\label{fig:HR-all}}
\end{figure}

Figure \ref{fig:HR-all} shows the H-R diagram of the 146 HBSs. The evolutionary tracks for the masses 1, 1.4, 2.2, 3, and 4 M$_{\odot}$ from the theoretical zero-age main sequence (ZAMS) for Z = 0.02 are produced with the stellar evolution code Modules for Experiments in Stellar Astrophysics (MESA; \citet{2011ApJS..192....3P, 2013ApJS..208....4P, 2015ApJS..220...15P, 2018ApJS..234...34P, 2019ApJS..243...10P, 2023ApJS..265...15J}) version 22.05.1 and MESASDK 21.4.1 \citep{2021zndo...5802444T}. We divide the HBSs with TEOs into two parts: the prominent TEO HBSs (orange pluses) and the non-prominent TEO HBSs (green pluses) (see Section \ref{sec:rst} for definitions). The eight HBSs reported in previous works (red pluses), i.e., KIC 8112039 (KOI-54) \citep{2011ApJS..197....4W}; KIC 3749404 \citep{2016MNRAS.463.1199H}; KIC 8164262 \citep{2017MNRAS.472L..25F}; KIC 4248941, KIC 5034333, KIC 8719324, and KIC 9016693 \citep{2020ApJ...888...95G}; KIC 11494130 \citep{2020ApJ...903..122C} also have prominent TEOs. As can be seen, the HBSs with prominent TEOs have higher surface effective temperatures, and this can provide observational evidence for \citet{2017MNRAS.472.1538F}'s suggestion that TEOs are more visible in hot stars with surface effective temperatures $T$ $\gtrsim$ 6500 K. In addition, these HBSs without TEOs (gray circles) are uniformly distributed. 

\subsection{Parameter Statistics} \label{subsec:relation}
\begin{figure}
	\gridline{
		\fig{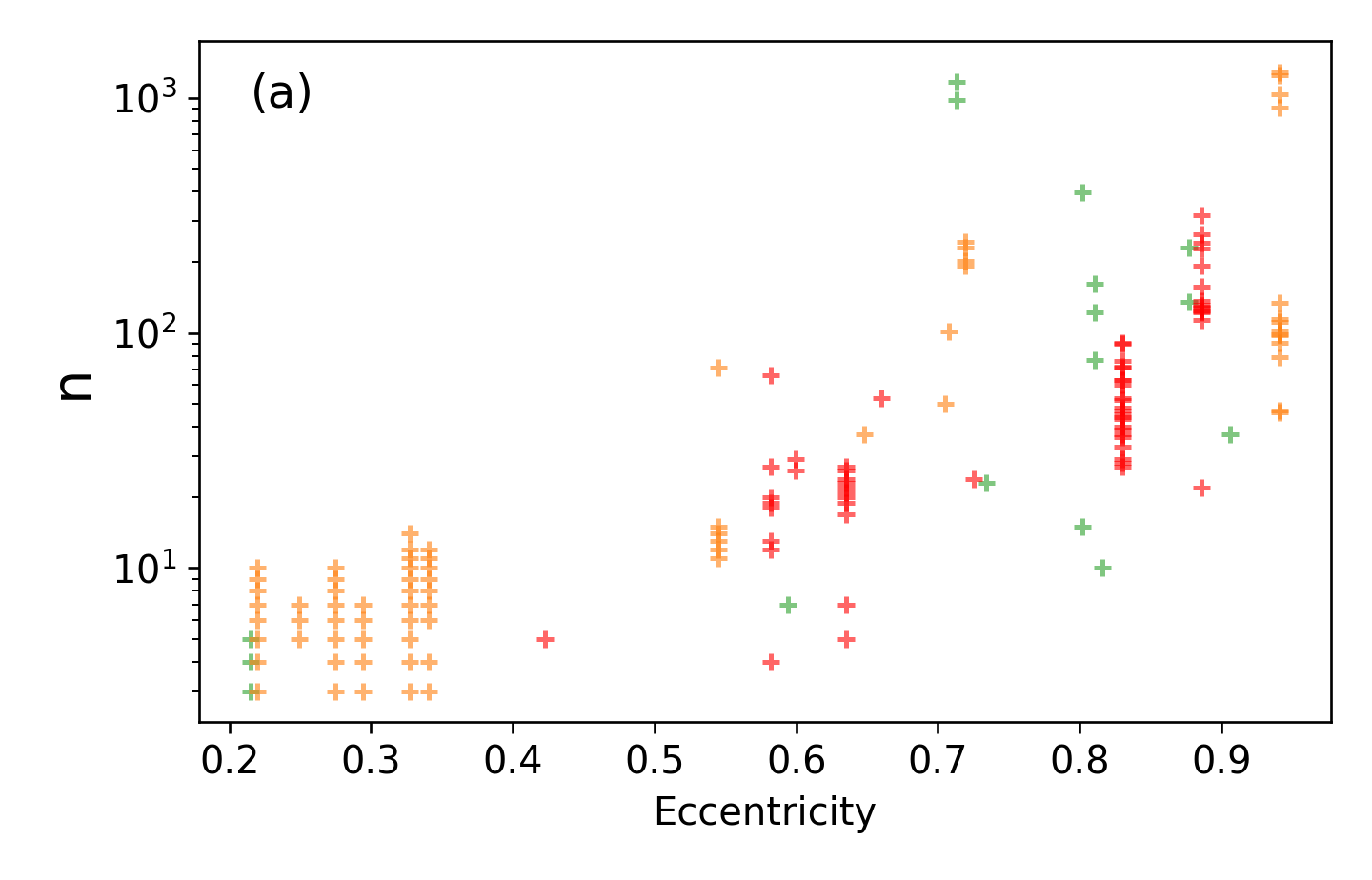}{0.5\textwidth}{}
		\fig{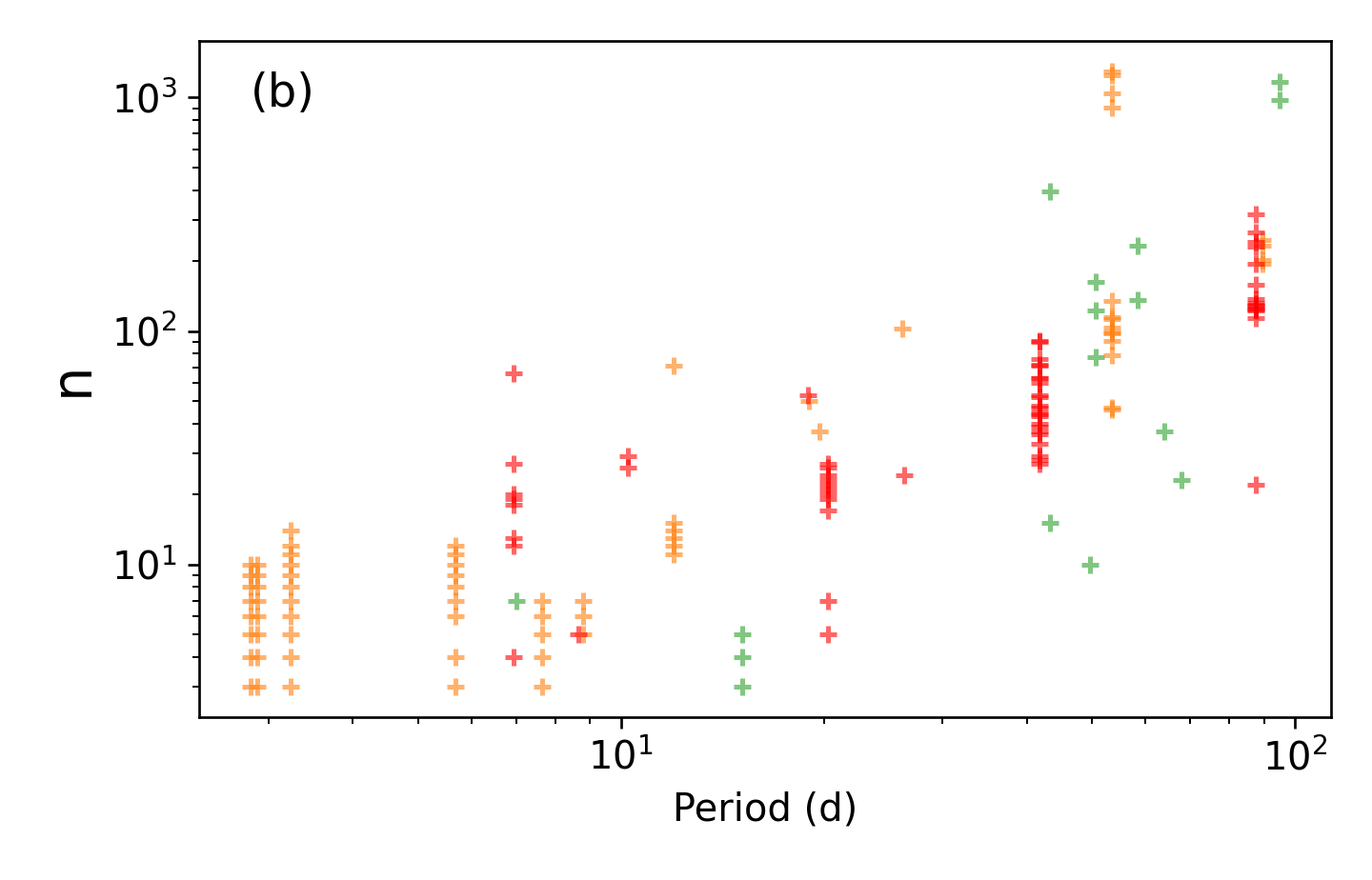}{0.5\textwidth}{}
	}
	\gridline{
		\fig{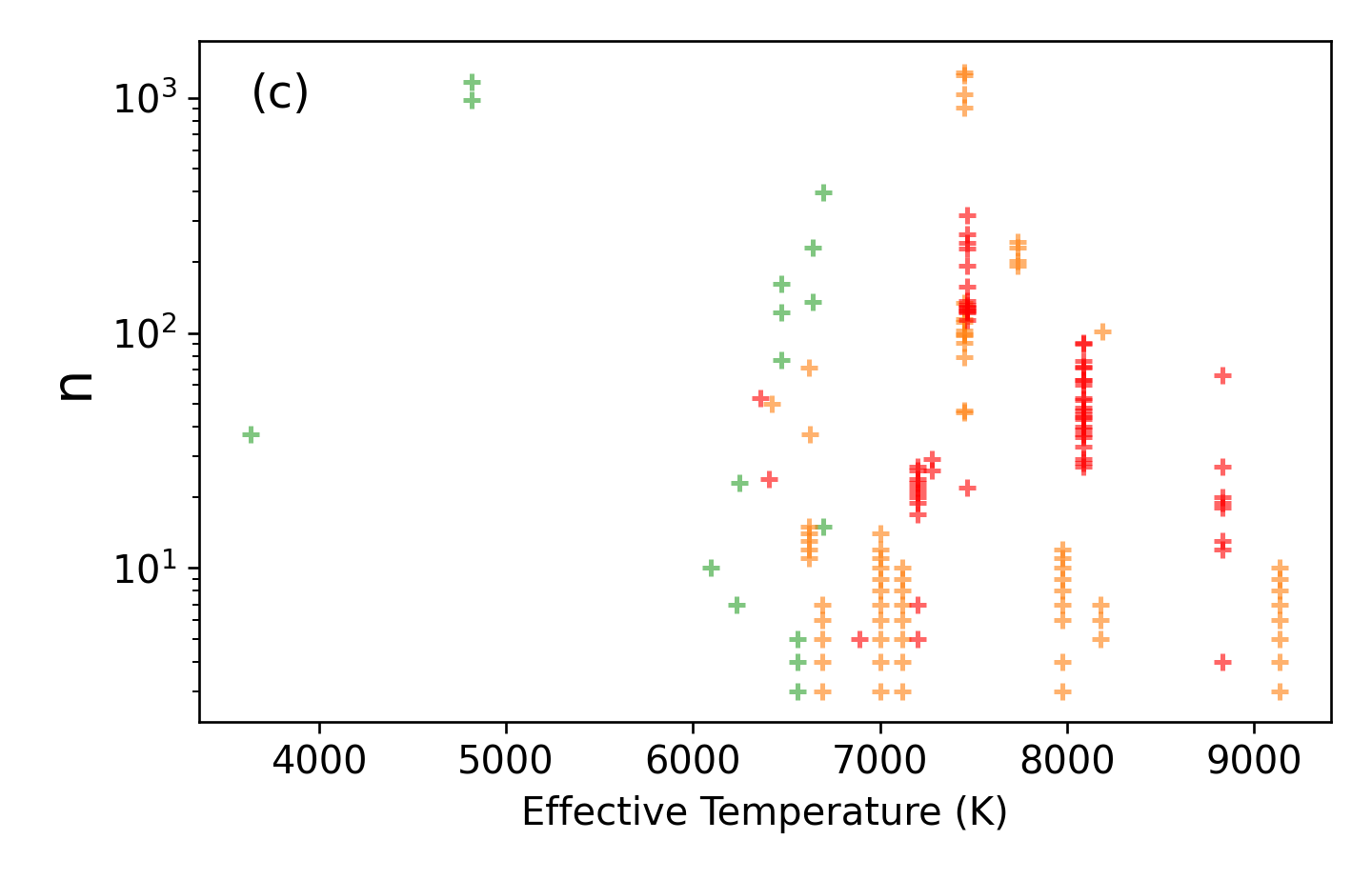}{0.5\textwidth}{}
		\fig{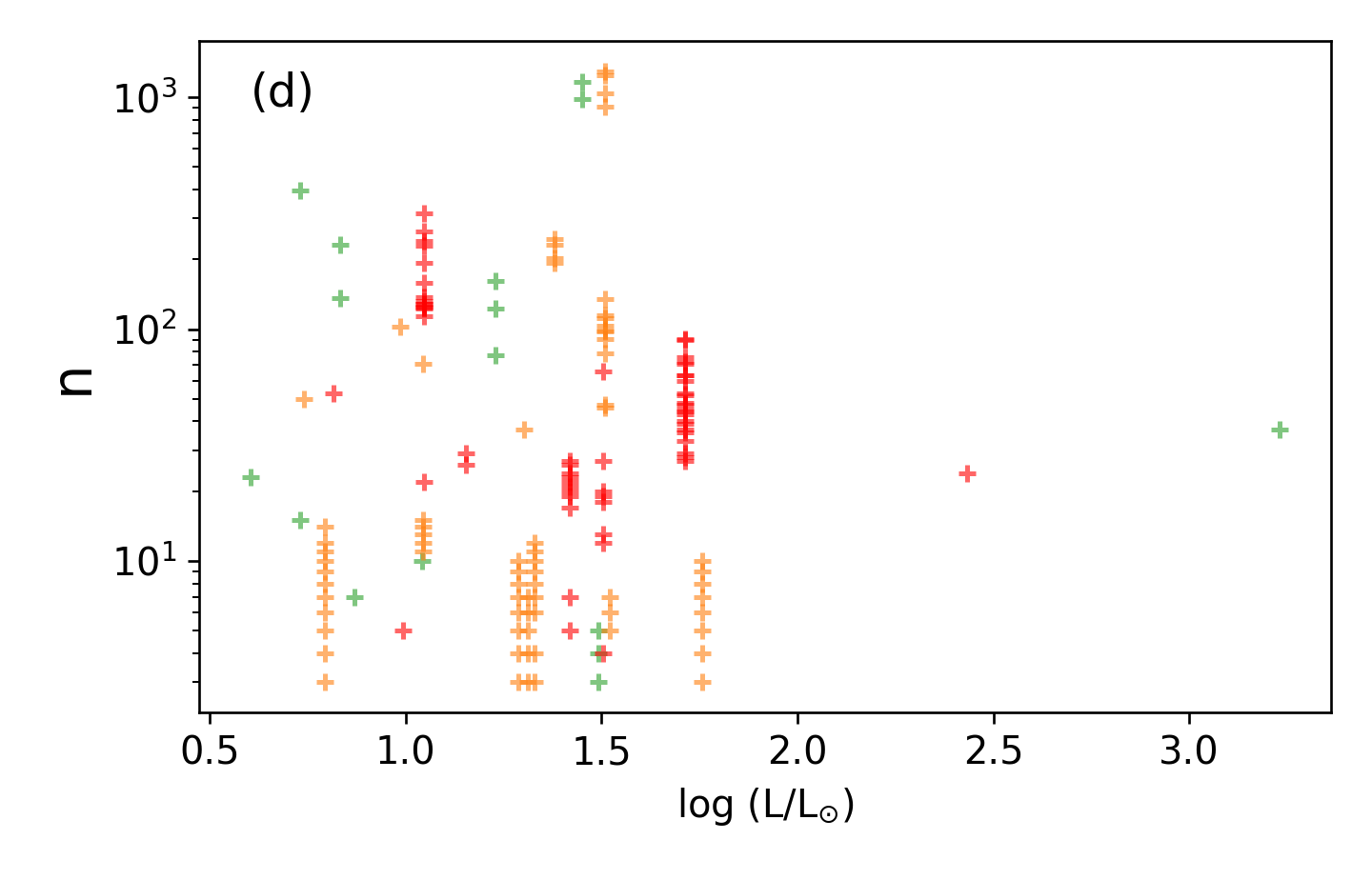}{0.5\textwidth}{}
	}
	\caption{(a) The relation between the eccentricities and the harmonic number $n$ of the TEOs. (b) The relation between the periods and $n$. (c) The relation between the effective temperature and $n$. (d) The relation between the luminosity and $n$. Orange and green pluses indicate the HBSs with prominent and non-prominent TEOs in Table \ref{tab:TEOs}, respectively. Red pluses indicate the TEOs in the other eight Kepler HBSs reported in previous works. 
		\label{fig:e_teos}}
\end{figure}

The four panels in Figure \ref{fig:e_teos} show the relation between the harmonic number $n$ of the TEOs and the eccentricity, period, temperature, and luminosity. They include not only the TEOs in Table \ref{tab:TEOs} (orange and green pluses for the HBSs with prominent and non-prominent TEOs, respectively), but also the TEOs in the eight previously reported HBSs mentioned in Section \ref{subsec:HR}. Note that the relations in panels (a) and (b) are positively correlated. The relation in panel (a) is evidence for the theoretical prediction of a positive correlation between eccentricity and $n$ \citep{2012MNRAS.421..983B, 2022ApJ...928..135W}. There is a positive correlation between period and eccentricity for these HBSs (see Figure 10 and Section 5.2 in \citetalias{2023ApJS..266...28L}); therefore, there is also a positive correlation between period and $n$ in panel (b). Panels (c) and (d) show no significant correlation. However, panel (c) also shows that TEOs are more likely to occur in hotter stars \citep{2017MNRAS.472.1538F}. Furthermore, we suggest that the luminosity distribution in panel (d) is due to observational bias, since these samples do not contain more stars with higher luminosity.

\begin{figure}
	\gridline{
		\fig{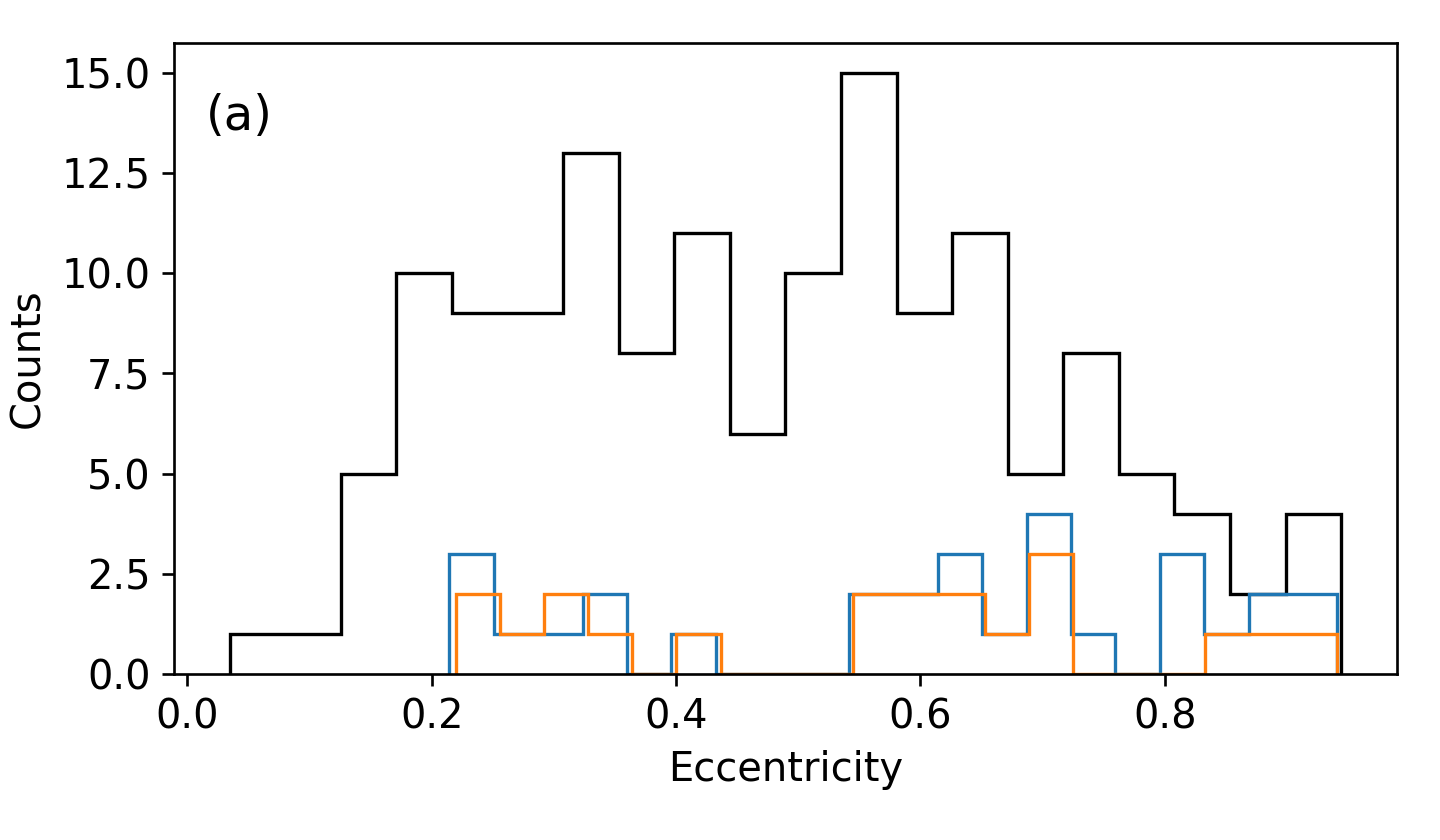}{0.5\textwidth}{}
		\fig{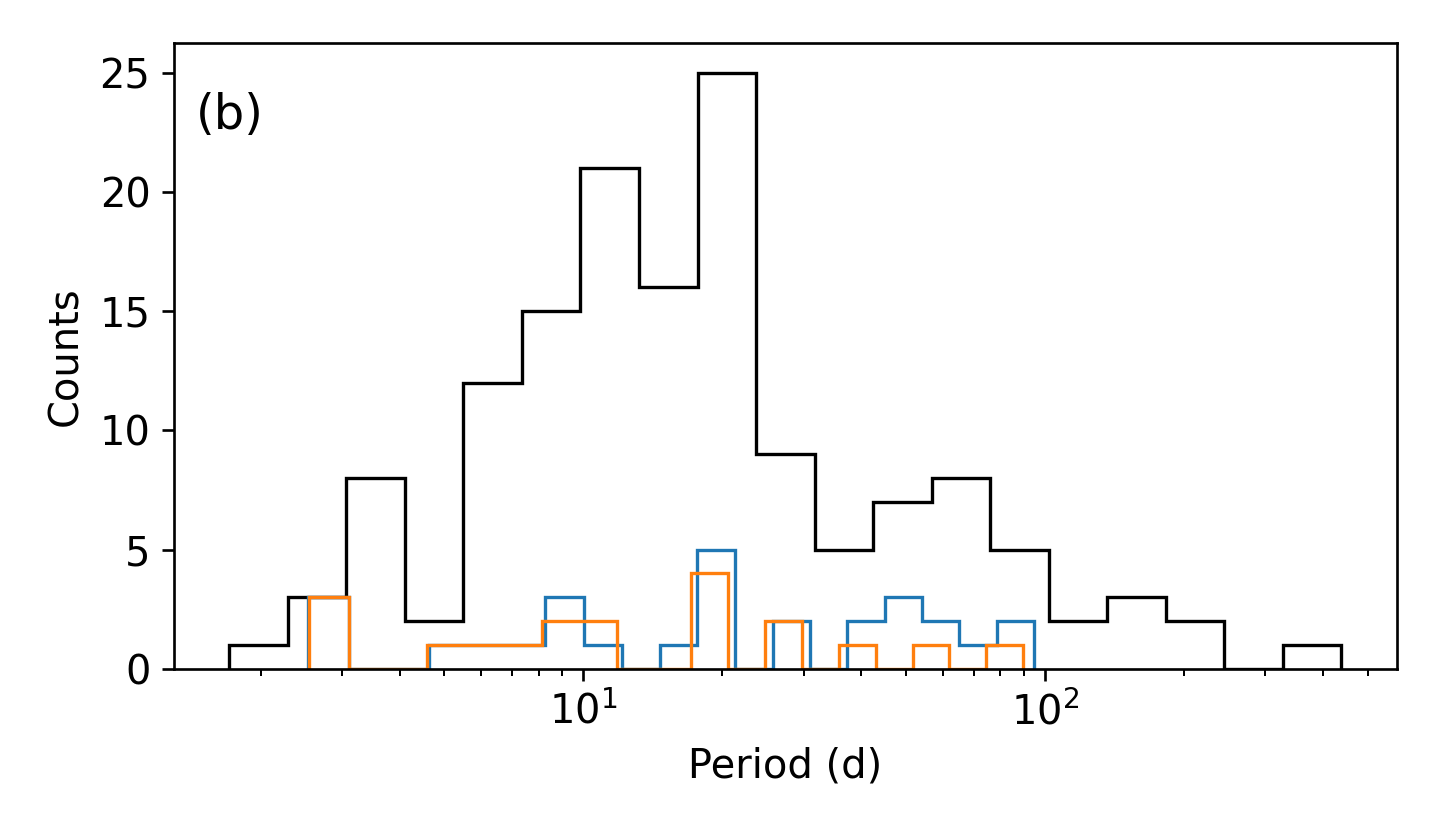}{0.5\textwidth}{}
	}
	\gridline{
		\fig{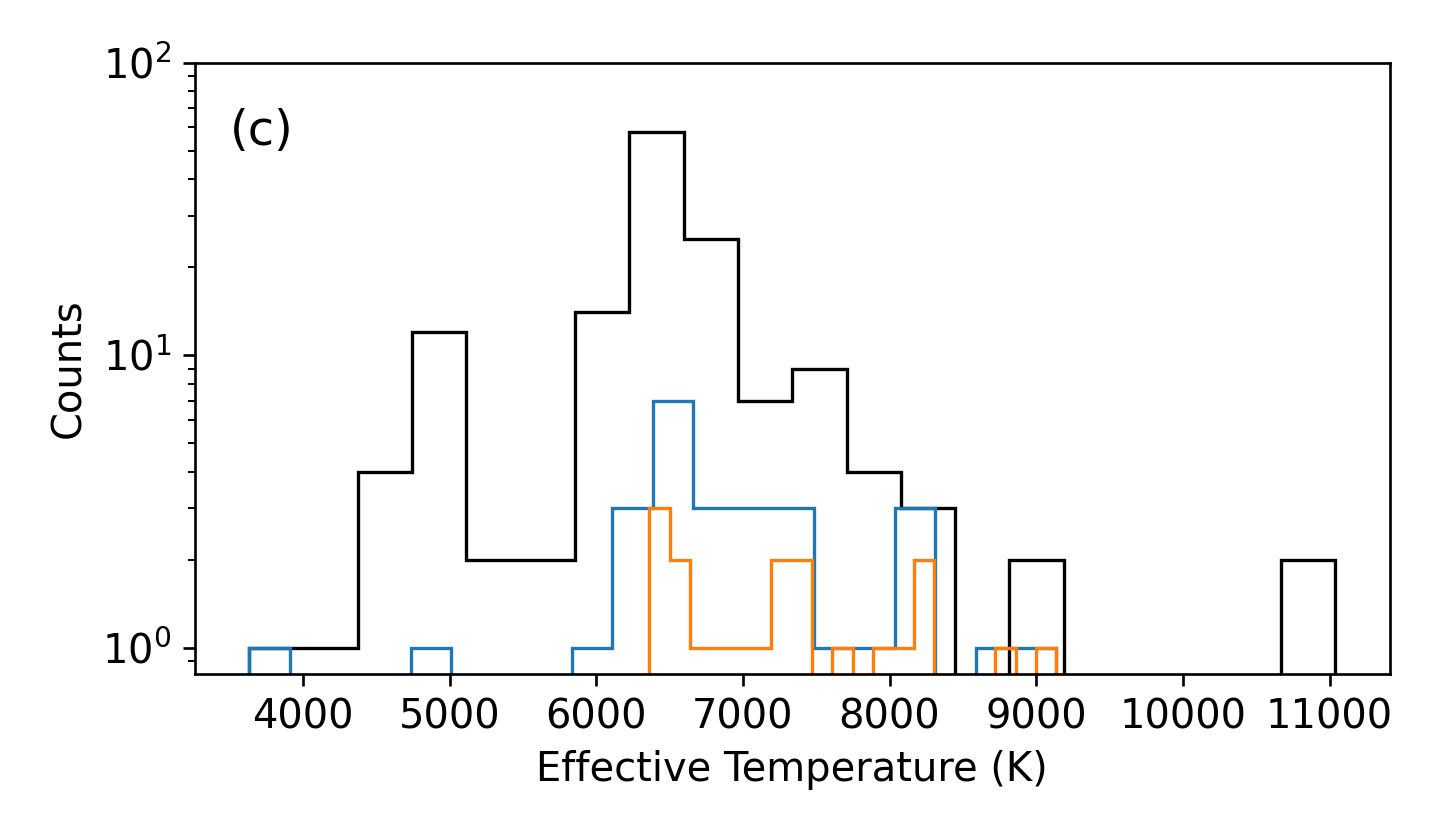}{0.5\textwidth}{}
		\fig{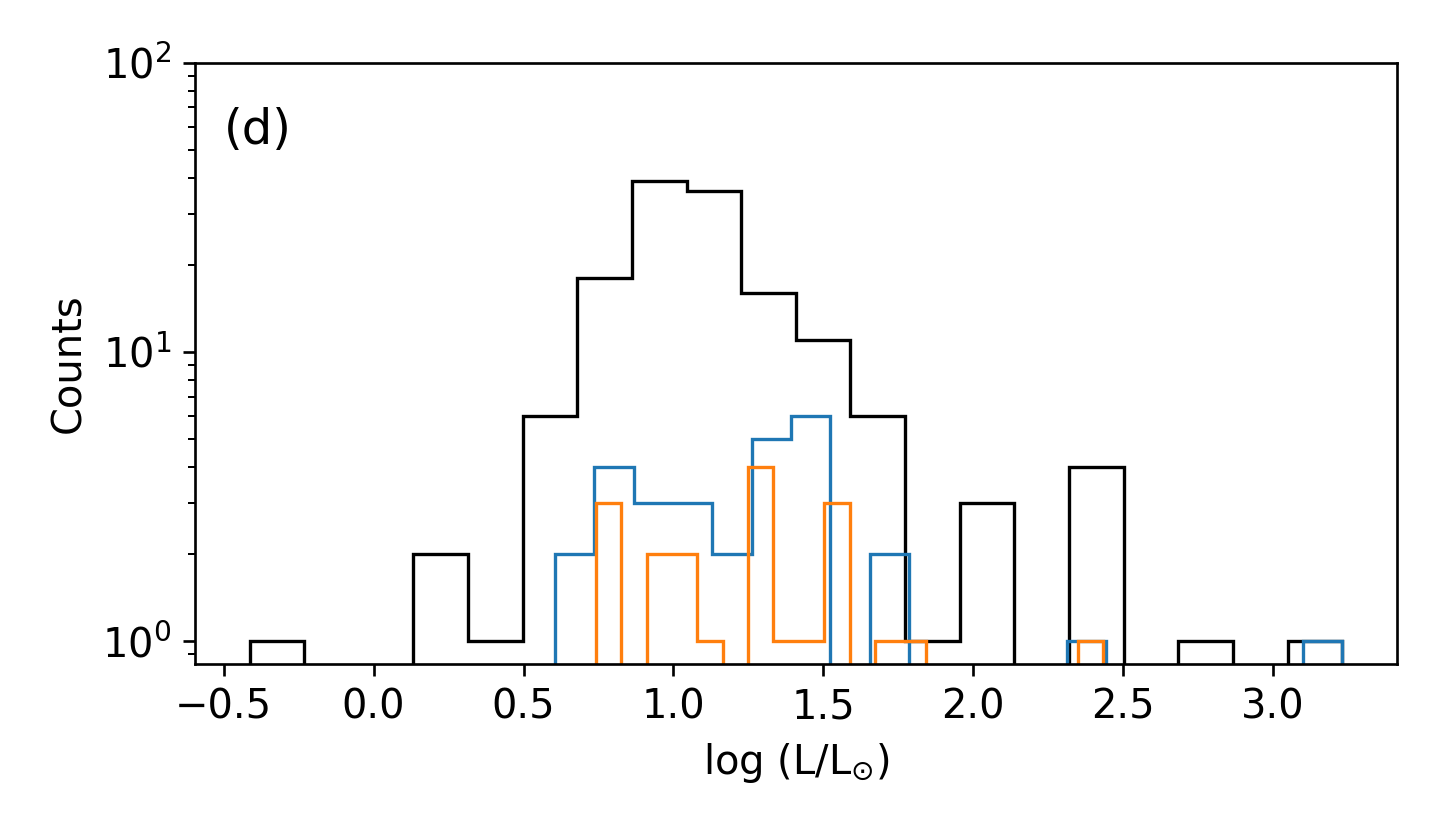}{0.5\textwidth}{}
	}
	\caption{Histograms of the eccentricity, period, effective temperature, and luminosity for the HBSs. Black curves represent all 146 HBSs. Blue curves represent the 21 HBSs with TEOs. Orange curves represent the 12 HBSs with prominent TEOs. 
		\label{fig:dis}}
\end{figure}

The four panels in Figure \ref{fig:dis} show the histograms of these parameters. In panels (a), (b), and (d), the distributions of these parameters of the HBSs with TEOs (blue curves) and HBSs with prominent TEOs (orange curves) appear relatively uniform. However, the absence of the prominent TEO in HBSs with orbital periods longer than 100 days in panel (b) may reveal that the relatively short period corresponds to strong tidal forces that can more easily form TEOs.

\section{Summary and Conclusions}\label{sec:conclusions}
In this work, we revisit the HBSs studied in \citetalias{2023ApJS..266...28L}, and examine their Fourier spectra. We compile a set of analytic procedures to find the harmonic TEOs and anharmonic TEOs, and newly analyze that 21 HBSs exhibit TEOs, 12 of which are prominent ($S/N \ge 10$). The relation between the orbital eccentricities and the harmonic number of the TEOs shows a positive correlation \citep{2012MNRAS.421..983B, 2022ApJ...928..135W}. The distribution of the HBSs in the H-R diagram (Figure \ref{fig:HR-all}) also shows that TEOs are more visible in hot stars with surface effective temperatures $T$ $\gtrsim$ 6500 K \citep{2017MNRAS.472.1538F}. The relation between the orbital periods and the harmonic number also shows a positive correlation in these HBSs. The histogram of the orbital period (Fig. \ref{fig:dis}- (b2)) may indicate that the relatively short period corresponds to strong tidal forces that can more easily form TEOs.

We present all the harmonic TEOs, and only the prominent anharmonic TEOs of these HBSs, but some HBSs have a significant number of anharmonic TEOs. More internal investigations are needed in future works. In addition, some of the systems marked as TEOs by visual inspection in \citetalias{2023ApJS..266...28L} should be reconsidered after a more precise Fourier spectral analysis. We suggest that some of them are self-excited or stochastically excited pulsating star candidates instead of TEO candidates, including KIC 4949194, KIC 7887124, KIC 7897952, KIC 8027591, KIC 8095275, KIC 9163796, KIC 9835416, KIC 9899216, KIC 10334122, KIC 11071278, KIC 11288684, and KIC 11409673, and leave a detailed study for future work.

\section{Acknowledgments}
This work is partly supported by the National Natural Science Foundation of China (Nos. 11933008 and 12103084), the Basic Research Project of Yunnan Province (Grant No. 202301AT070352), and the Yunnan Revitalization Talent Support Program. We would like to thank \citet{2016AJ....151...68K} for their excellent work and for sharing their results. Funding for the Kepler mission is provided by the NASA Science Mission directorate. The Guoshoujing Telescope (the Large Sky Area Multi-Object Fiber Spectroscopic Telescope LAMOST) is a National Major Scientific Project built by the Chinese Academy of Sciences. The Gaia Survey is a cornerstone mission of the European Space Agency. We thank the Kepler, LAMOST, and Gaia teams for their support and hard work. We are grateful to the anonymous referee for his or her constructive comments that improved this manuscript.

\software{
	FNPEAKS (Z. Koo{\l}aczkowski, W. Hebisch, G. Kopacki), 
		MESA \citep{2011ApJS..192....3P, 2013ApJS..208....4P, 2015ApJS..220...15P, 2018ApJS..234...34P, 2019ApJS..243...10P, 2023ApJS..265...15J},
		MESASDK \citep{2021zndo...5802444T}.
}




\bibliography{sample631}
\bibliographystyle{aasjournal}

\appendix
\setcounter{figure}{0}
\renewcommand{\thefigure}{A\arabic{figure}}
\section{The Analytic Procedure for the Other 19 HBSs} \label{sec:apdx}

\begin{figure}
	\centering
	\includegraphics[width=0.75\textwidth]{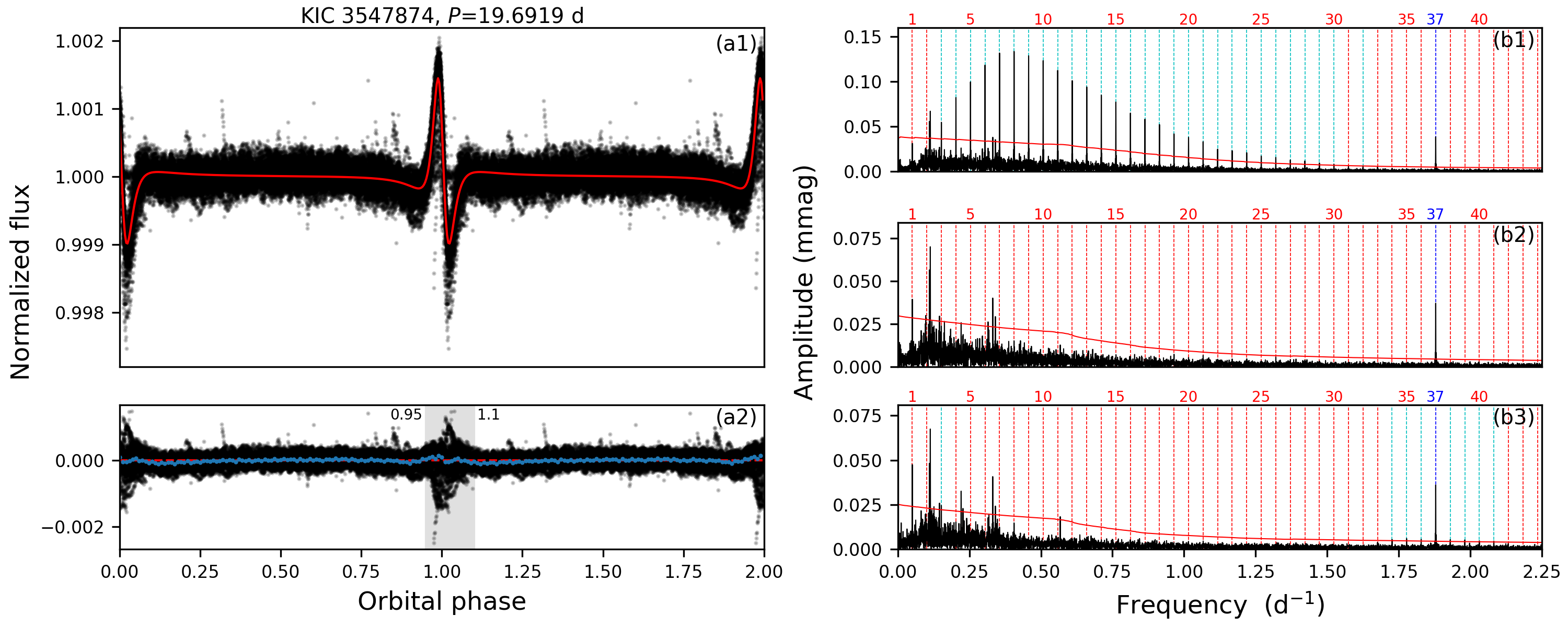}
	\caption{The analytic procedure for KIC 3547874. The TEO candidate is the $n$ = 37 harmonic.
		\label{fig:3547874-ft}}
\end{figure}
\begin{figure}
	\centering
	\includegraphics[width=0.75\textwidth]{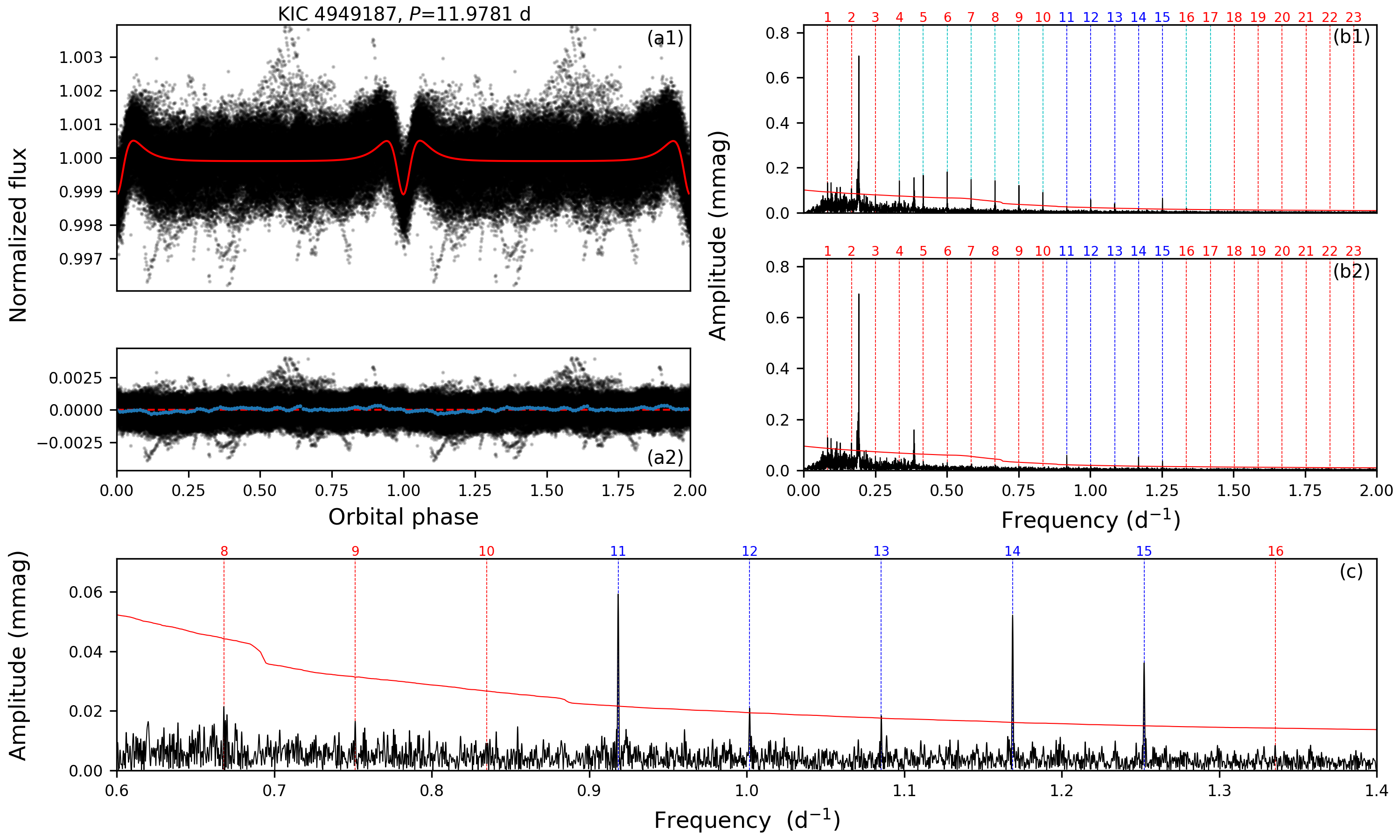}
	\caption{The analytic procedure for KIC 4949187. Panel (c) zooms in on panel (b2). The TEO candidates are the $n$ = 11, 14, 15, 12, and 13 harmonics.
		\label{fig:4949187-ft}}
\end{figure}
\begin{figure}
	\centering
	\includegraphics[width=0.75\textwidth]{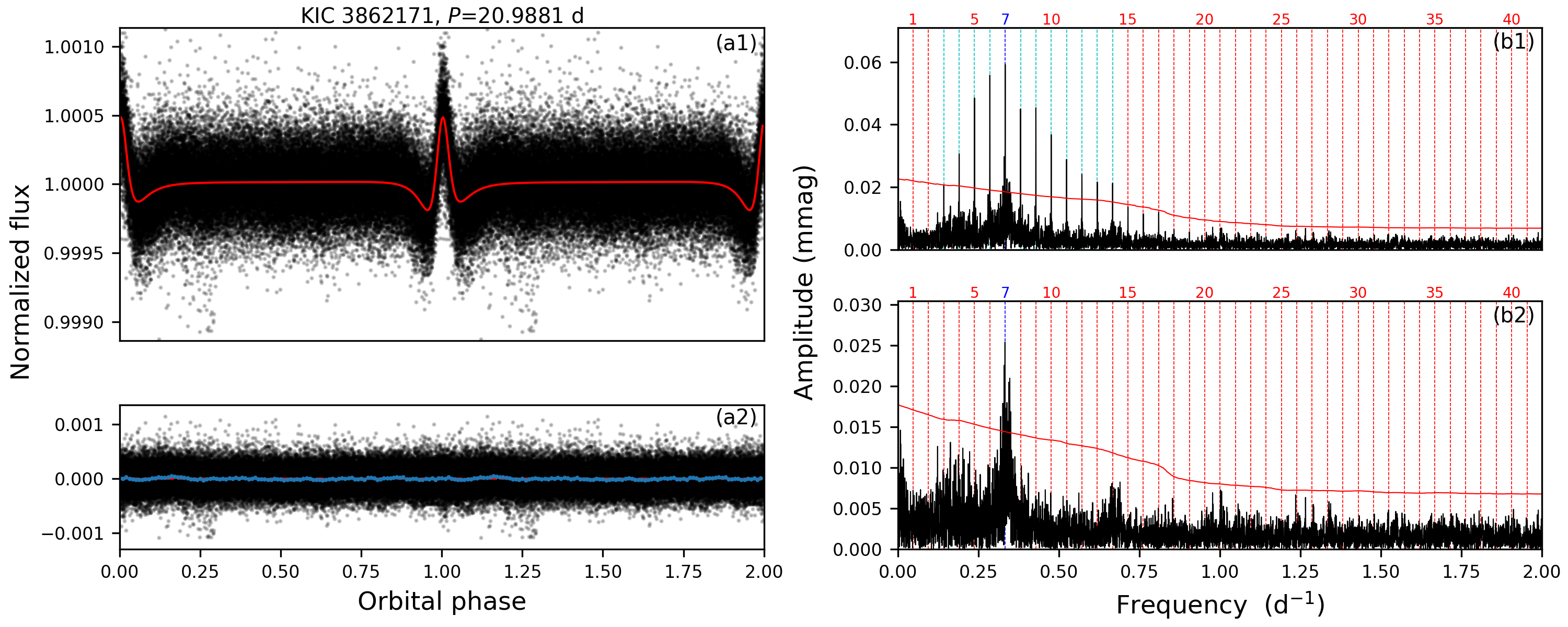}
	\caption{The analytic procedure for KIC 3862171. The TEO candidate is the $n$ = 7 harmonic.
		\label{fig:3862171-ft}}
\end{figure}
\begin{figure}
	\centering
	\includegraphics[width=0.75\textwidth]{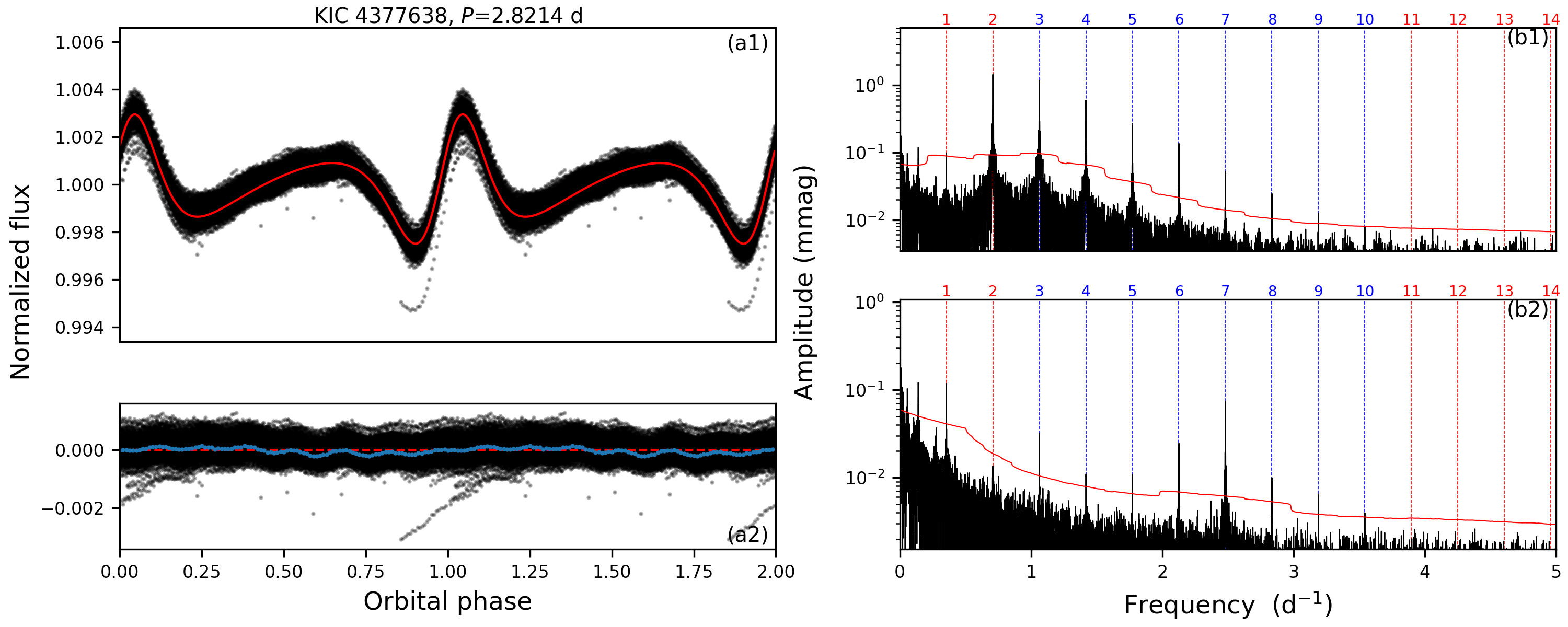}
	\caption{The analytic procedure for KIC 4377638. The TEO candidates are the $n$ = 7, 3, 6, 4, 5, 8, 9, and 10 harmonics. Although other harmonics may be dominated by imperfect fit, the $n$ = 7 orbital harmonic stands out clearly and cannot be explained by the imperfect removal. 
		\label{fig:4377638-ft}}
\end{figure}

\begin{figure}
	\centering
	\includegraphics[width=0.75\textwidth]{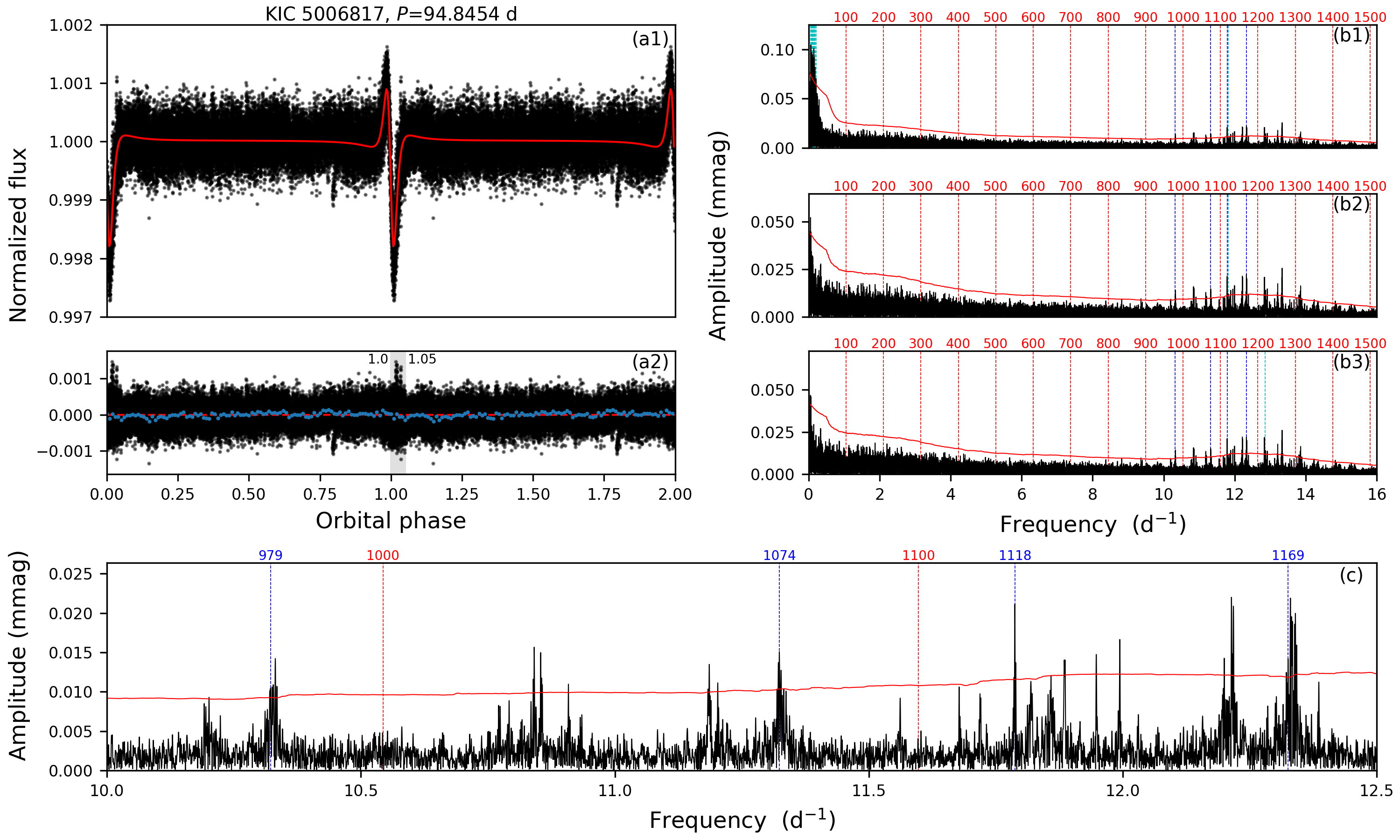}
	\caption{The analytic procedure for KIC 5006817. Panel (c) zooms in on panel (b3). The TEO candidates are the $n$ = 1118, 1074, 1169, and 979 harmonics.
		\label{fig:5006817-ft}}
\end{figure}
\begin{figure}
	\centering
	\includegraphics[width=0.75\textwidth]{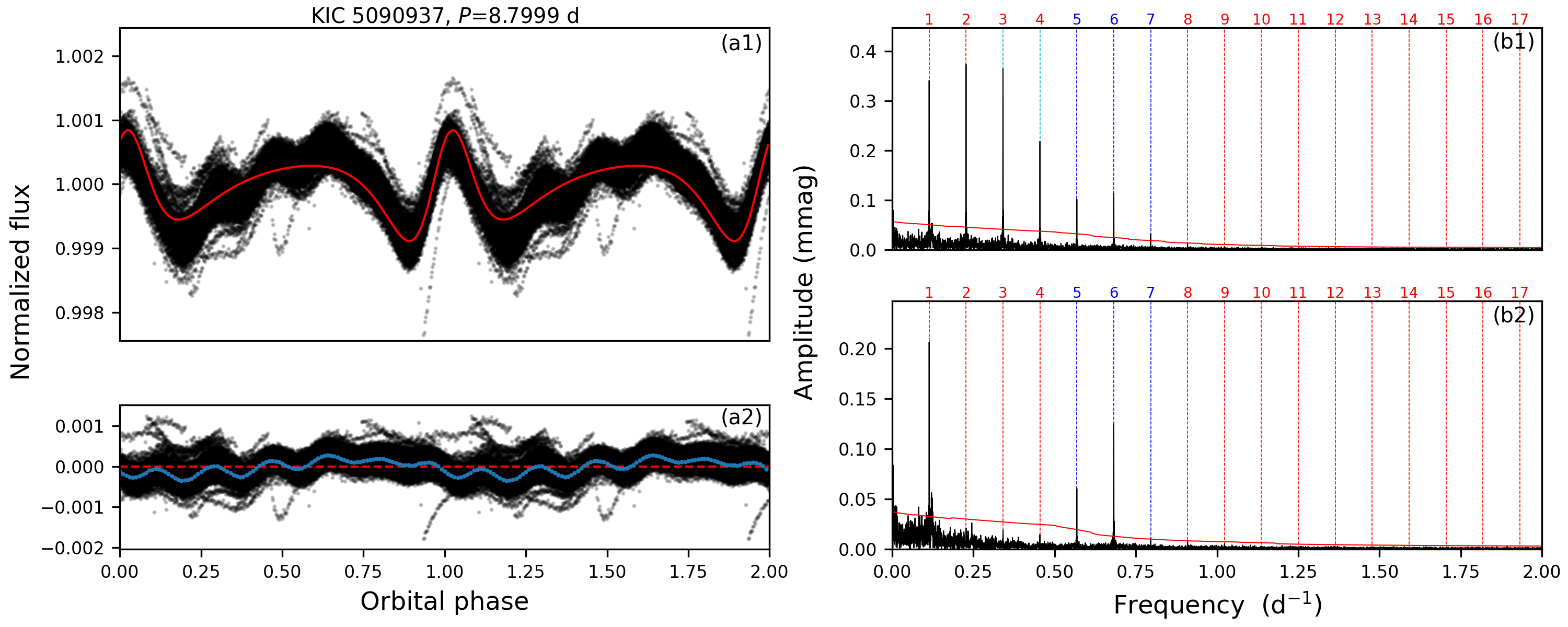}
	\caption{The analytic procedure for KIC 5090937. The TEO candidates are the $n$ = 6, 5, and 7 harmonics.
		\label{fig:5090937-ft}}
\end{figure}
\begin{figure}
	\centering
	\includegraphics[width=0.75\textwidth]{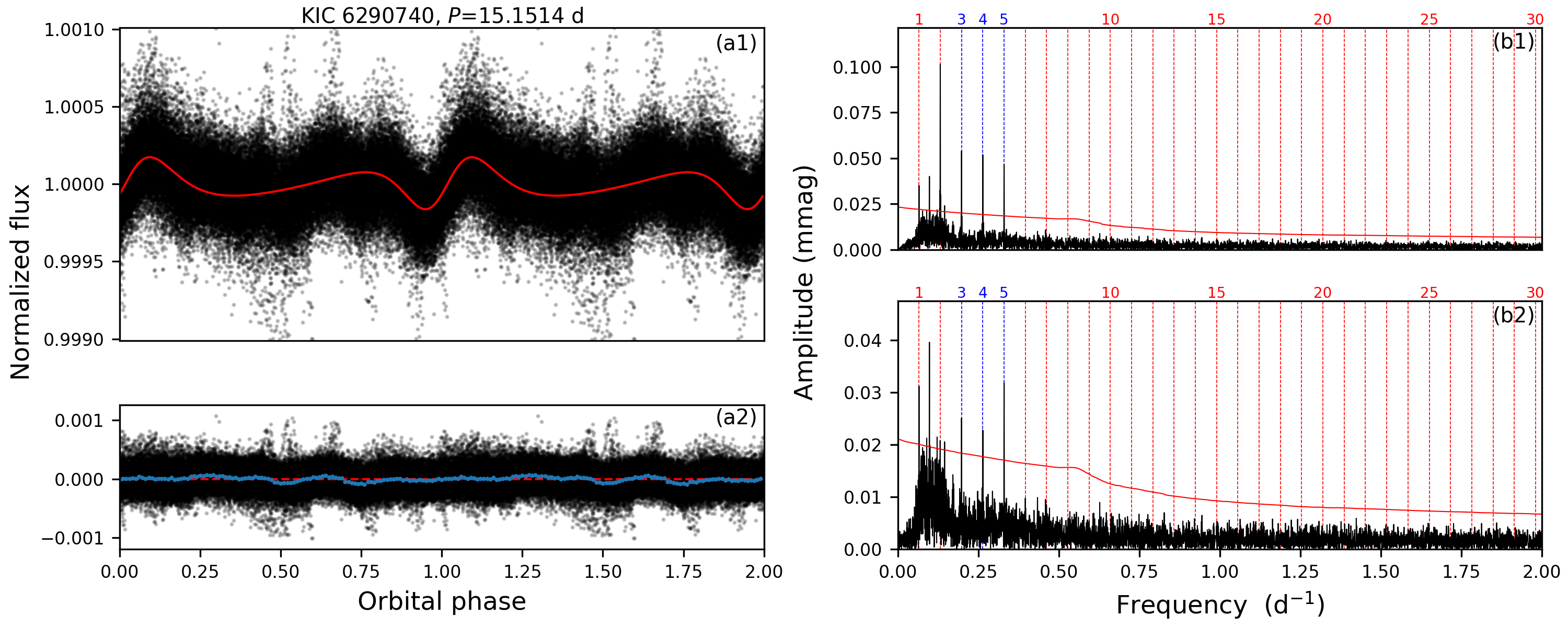}
	\caption{The analytic procedure for KIC 6290740. The TEO candidates are the $n$ = 5, 3, and 4 harmonics. Although the $n$ = 3, 4 harmonics may be dominated by imperfect fit, the $n$ = 5 orbital harmonic stands out clearly and cannot be explained by the imperfect removal. 
		\label{fig:6290740-ft}}
\end{figure}

\begin{figure}
	\centering
	\includegraphics[width=0.75\textwidth]{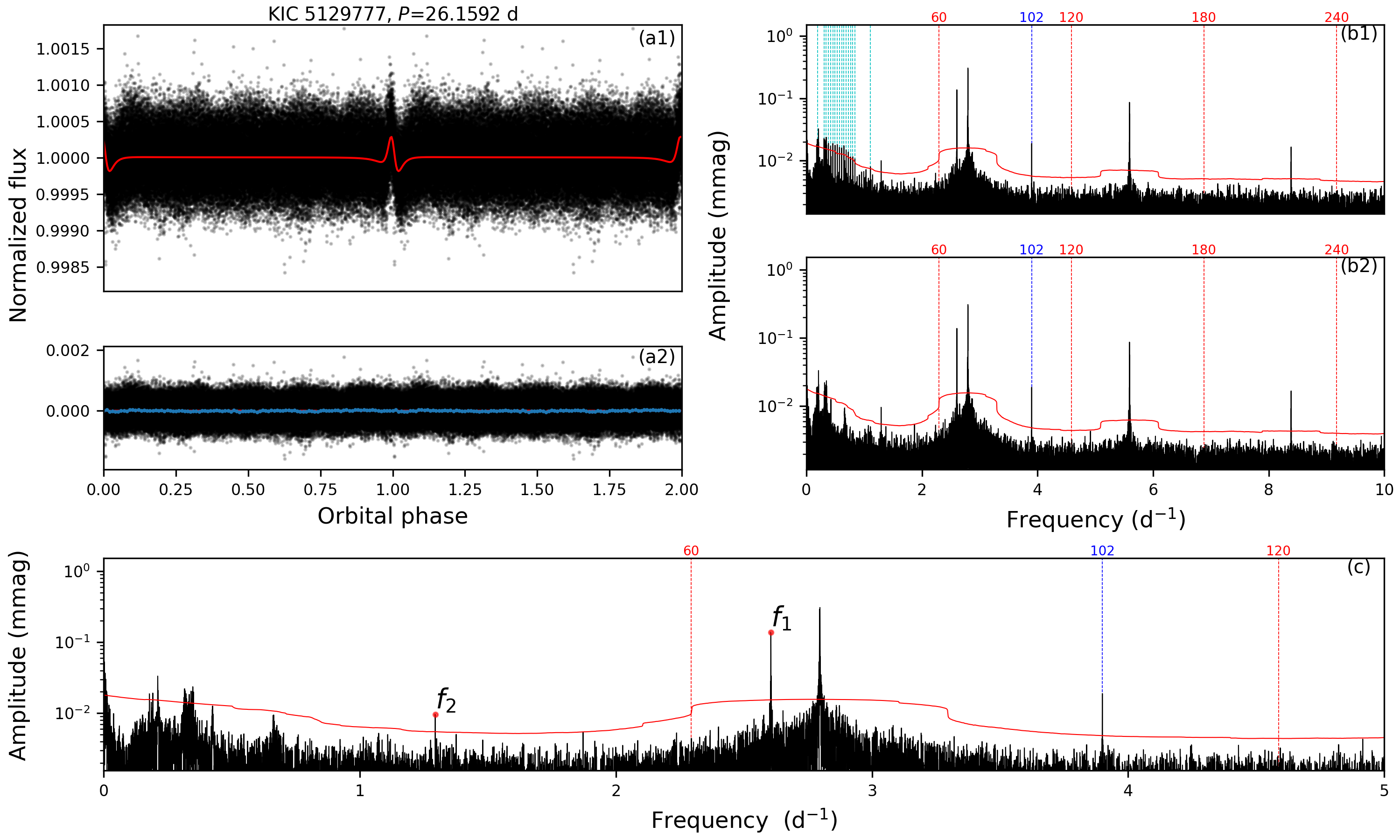}
	\caption{The analytic procedure for KIC 5129777. Panel (c) zooms in on panel (b2). The TEO candidate is the $n$ = 102 harmonic. The $f_1$ and $f_2$ are second-order modes coupling and satisfy $f_1$+$f_2$=102$f_{orb}$.
		\label{fig:5129777-ft}}
\end{figure}
\begin{figure}
	\centering
	\includegraphics[width=0.75\textwidth]{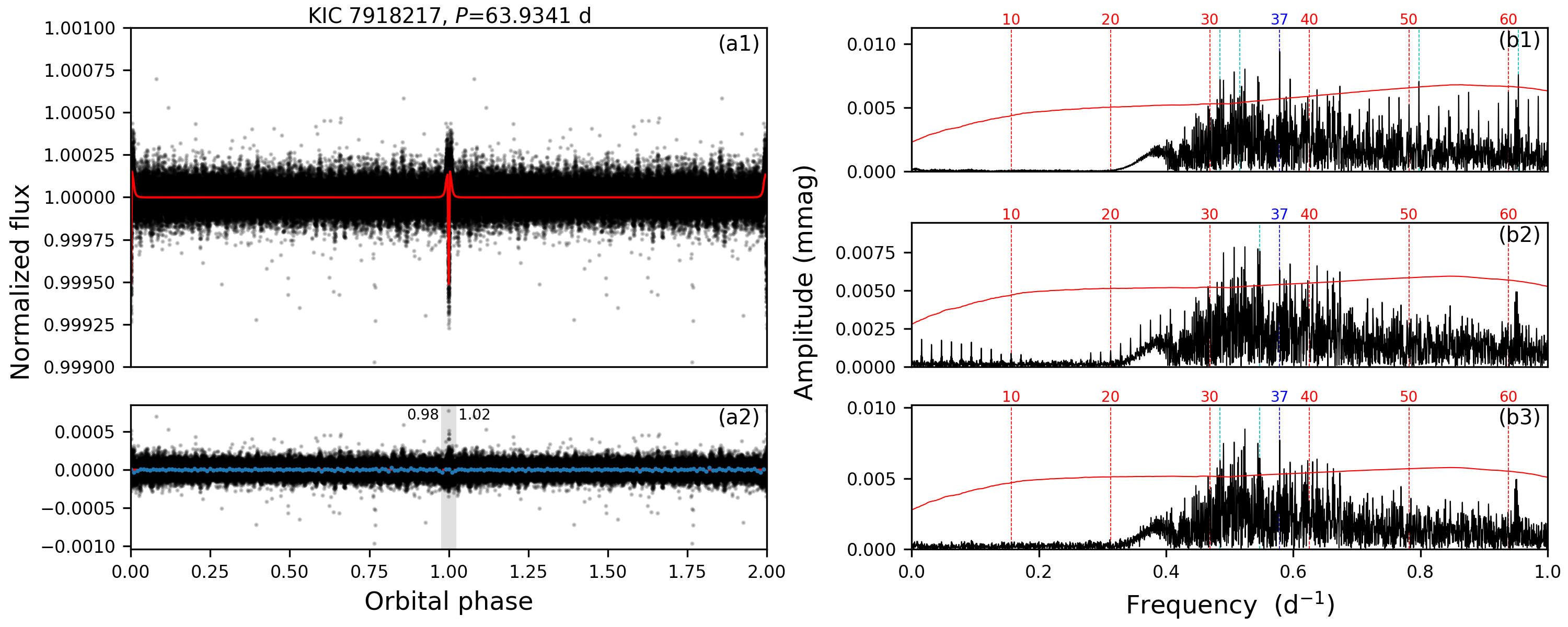}
	\caption{The analytic procedure for KIC 7918217. The TEO candidate is the $n$ = 37 harmonic.
		\label{fig:7918217-ft}}
\end{figure}
\begin{figure}
	\centering
	\includegraphics[width=0.75\textwidth]{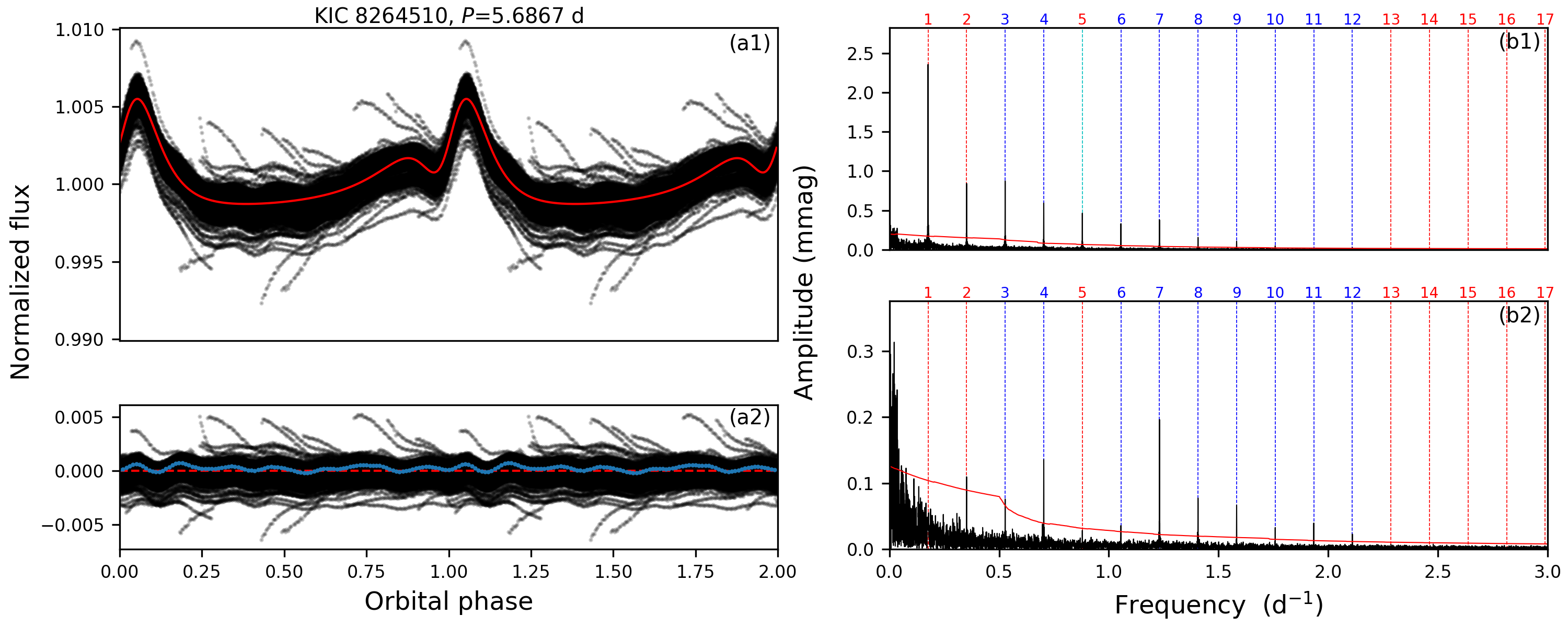}
	\caption{The analytic procedure for KIC 8264510. The TEO candidates are the $n$ = 7, 4, 8, 3, 9, 11, 6, 10, and 12 harmonics. The $n$ = 7, 4, 8, 9, and 11 harmonics stand out clearly.
		\label{fig:8264510-ft}}
\end{figure}

\begin{figure}
	\centering
	\includegraphics[width=0.75\textwidth]{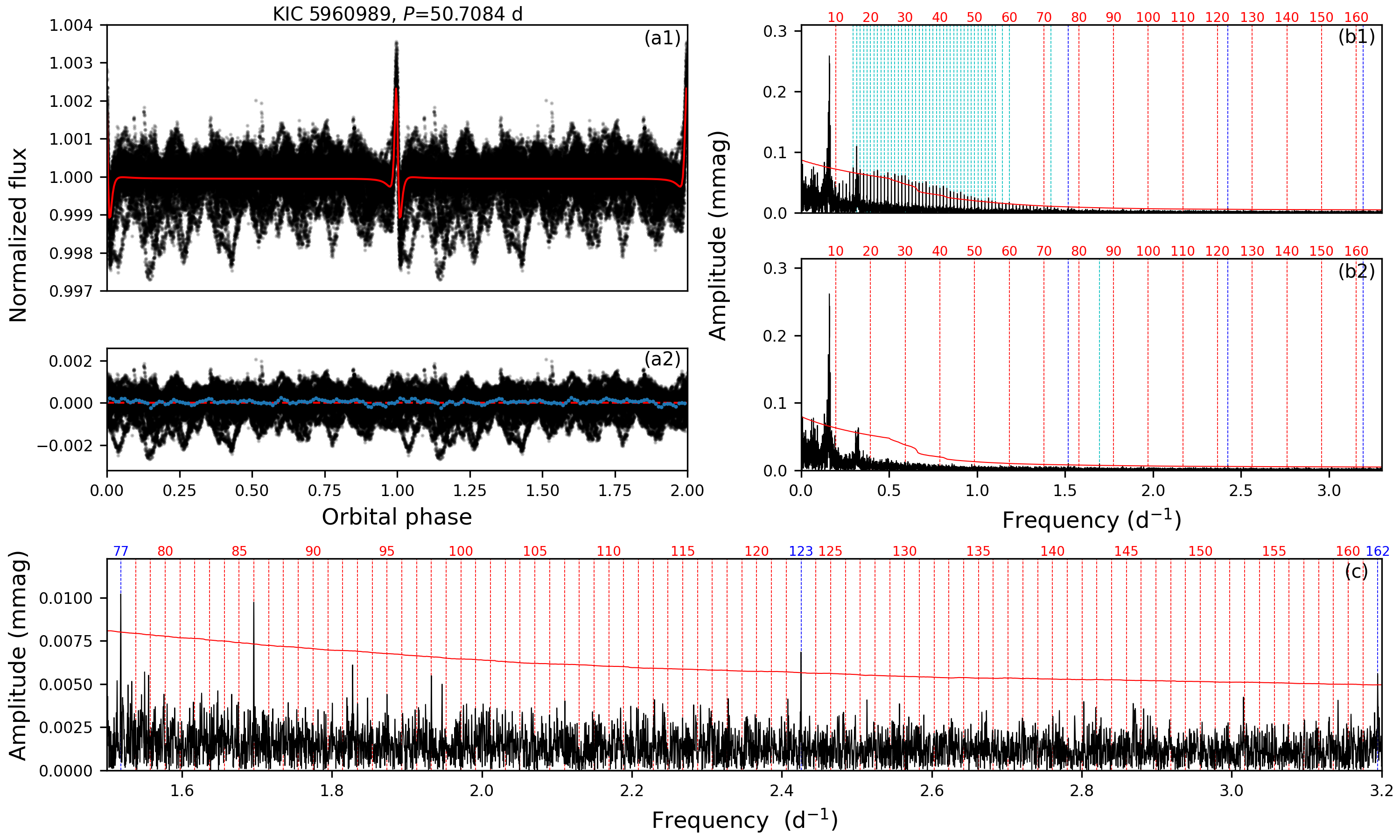}
	\caption{The analytic procedure for KIC 5960989. Panel (c) zooms in on panel (b2). The TEO candidates are the $n$ = 77, 123, and 162 harmonics.
		\label{fig:5960989-ft}}
\end{figure}

\begin{figure}
	\centering
	\includegraphics[width=0.75\textwidth]{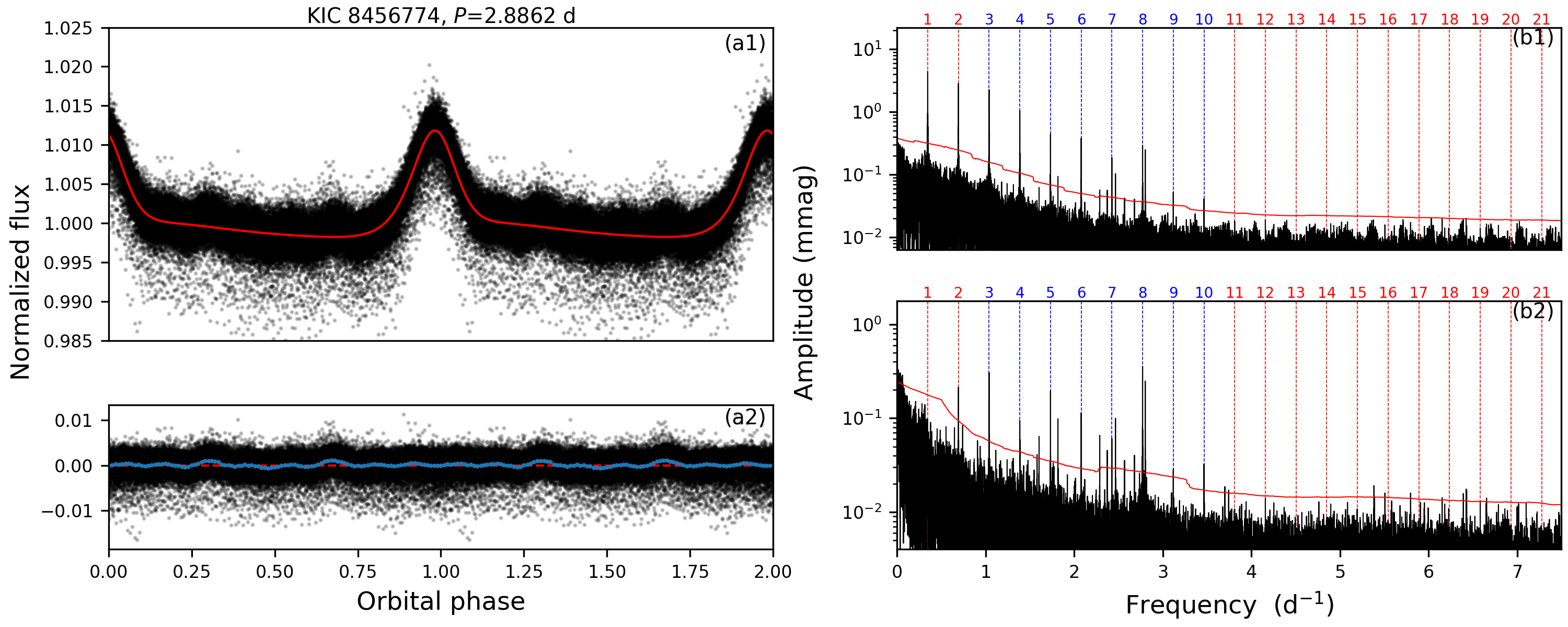}
	\caption{The analytic procedure for KIC 8456774. The TEO candidates are the $n$ = 8, 3, 5, 6, 4, 7, 10, and 9 harmonics. The $n$ = 8, 3, 5, and 6 harmonics stand out clearly.
		\label{fig:8456774-ft}}
\end{figure}
\begin{figure}
	\centering
	\includegraphics[width=0.75\textwidth]{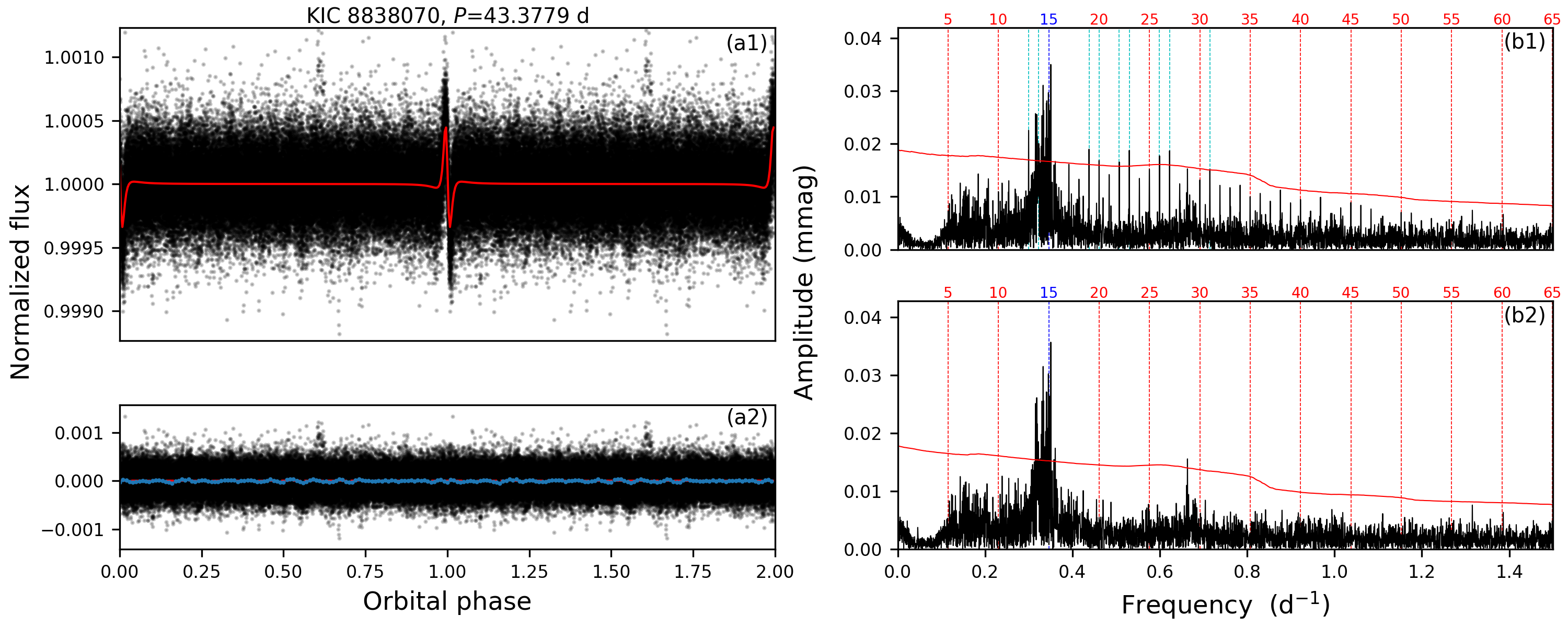}
	\caption{The analytic procedure for KIC 8838070. The TEO candidate is the $n$ = 15 harmonic.
		\label{fig:8838070-ft}}
\end{figure}
\begin{figure}
	\centering
	\includegraphics[width=0.75\textwidth]{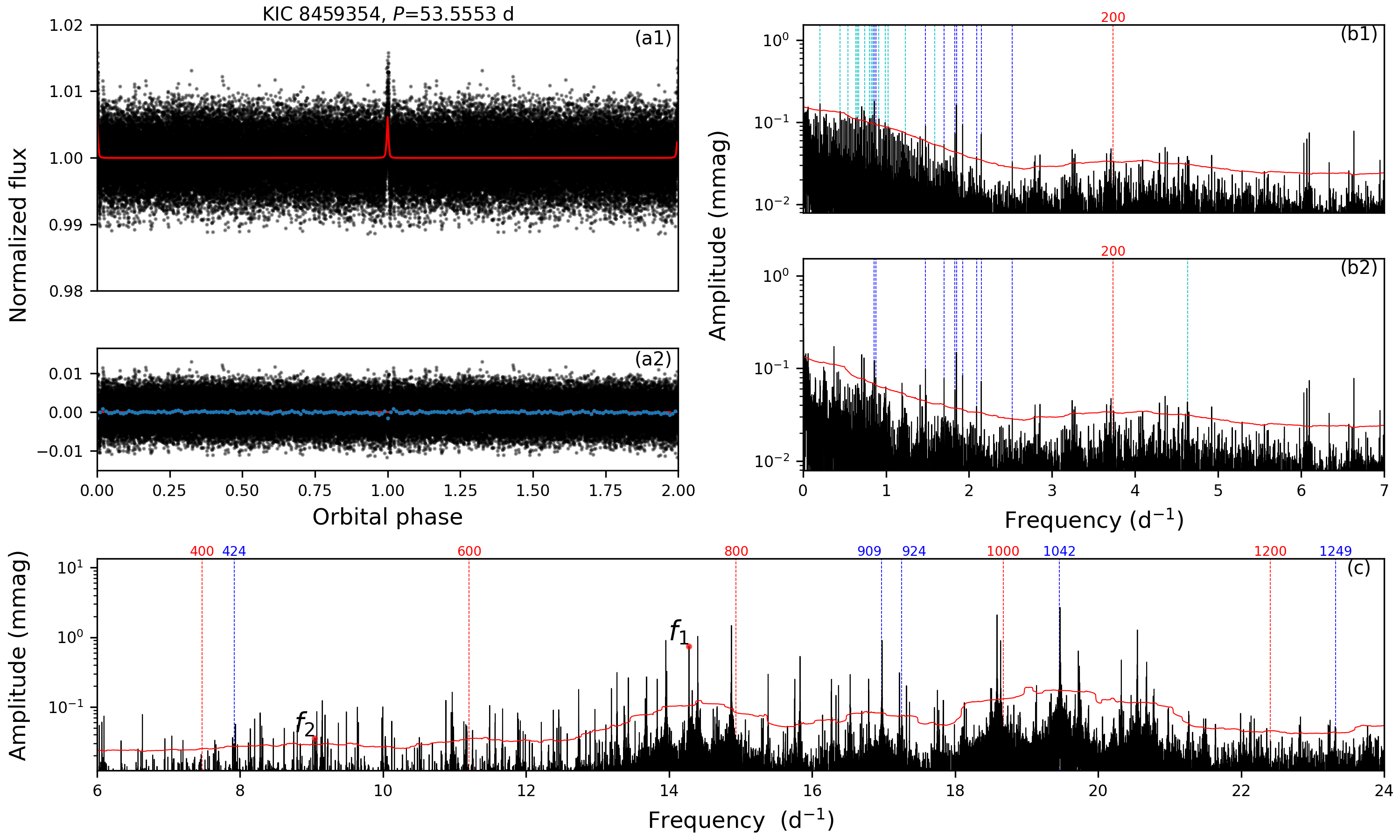}
	\caption{The analytic procedure for KIC 8459354. Panel (c) zooms in on panel (b2). The TEO candidates are the $n$ = 99, 46, 79, 47, 91, 115, 98, 112, 135, 424, 909, 924, 1042, and 1249 harmonics (lower harmonics are shown in panel (b2)). The $f_1$ and $f_2$ are second-order modes coupling and satisfy $f_1$+$f_2$=1249$f_{orb}$.
		\label{fig:8459354-ft}}
\end{figure}
\begin{figure}
	\centering
	\includegraphics[width=0.75\textwidth]{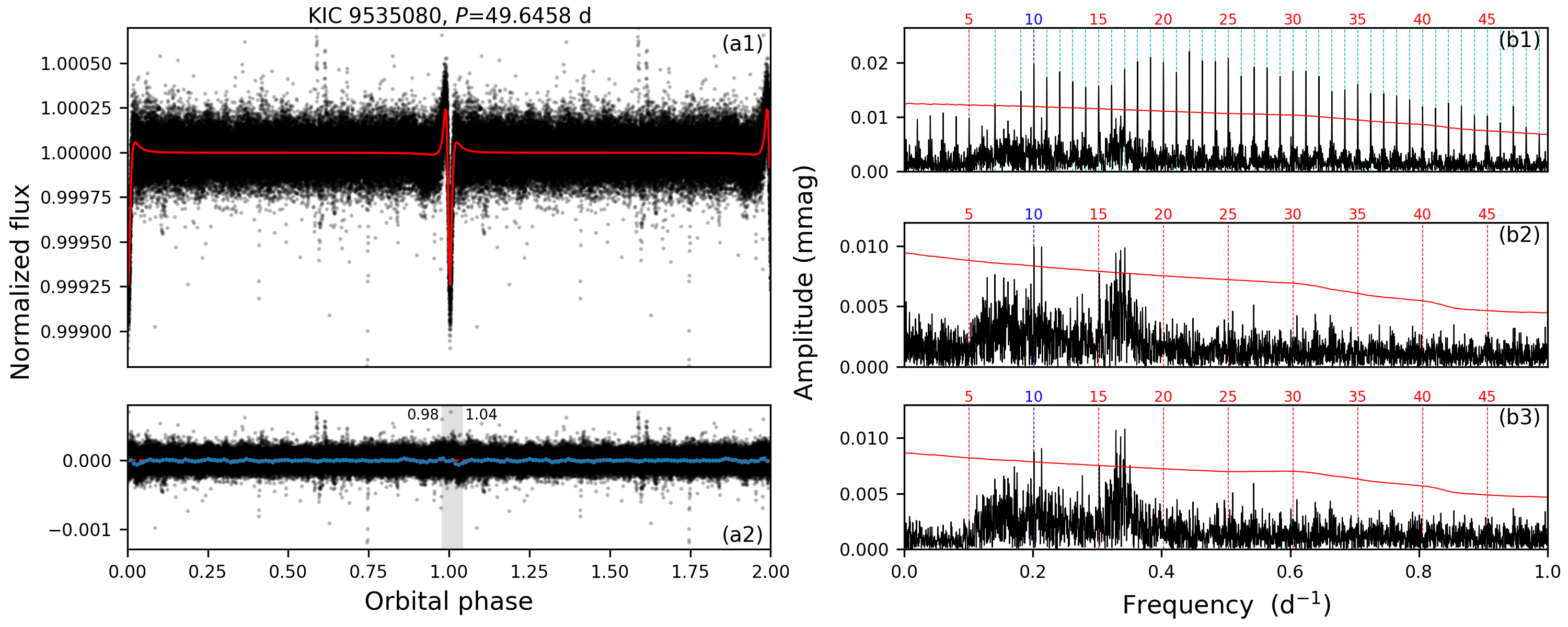}
	\caption{The analytic procedure for KIC 9535080. The TEO candidate is the $n$ = 10 harmonic.
		\label{fig:9535080-ft}}
\end{figure}
\begin{figure}
	\centering
	\includegraphics[width=0.75\textwidth]{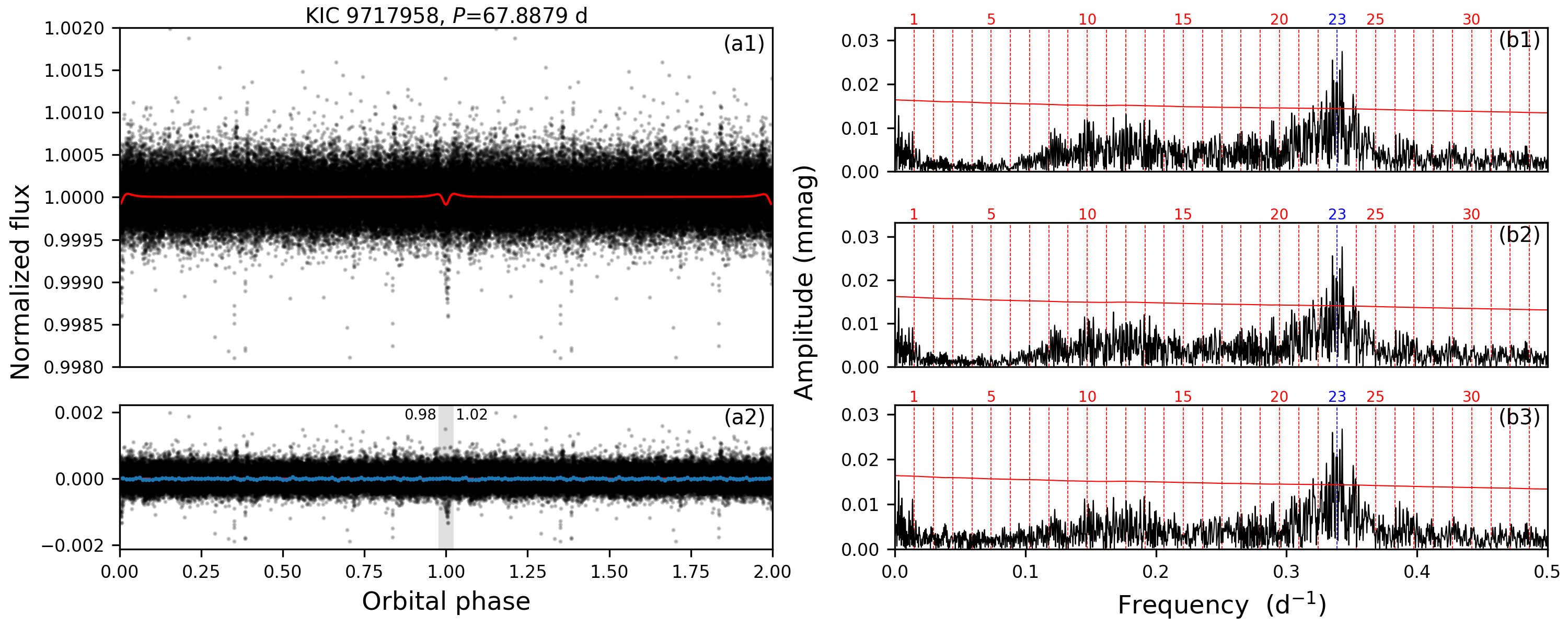}
	\caption{The analytic procedure for KIC 9717958. The TEO candidate is the $n$ = 23 harmonic.
		\label{fig:9717958-ft}}
\end{figure}
\begin{figure}
	\centering
	\includegraphics[width=0.75\textwidth]{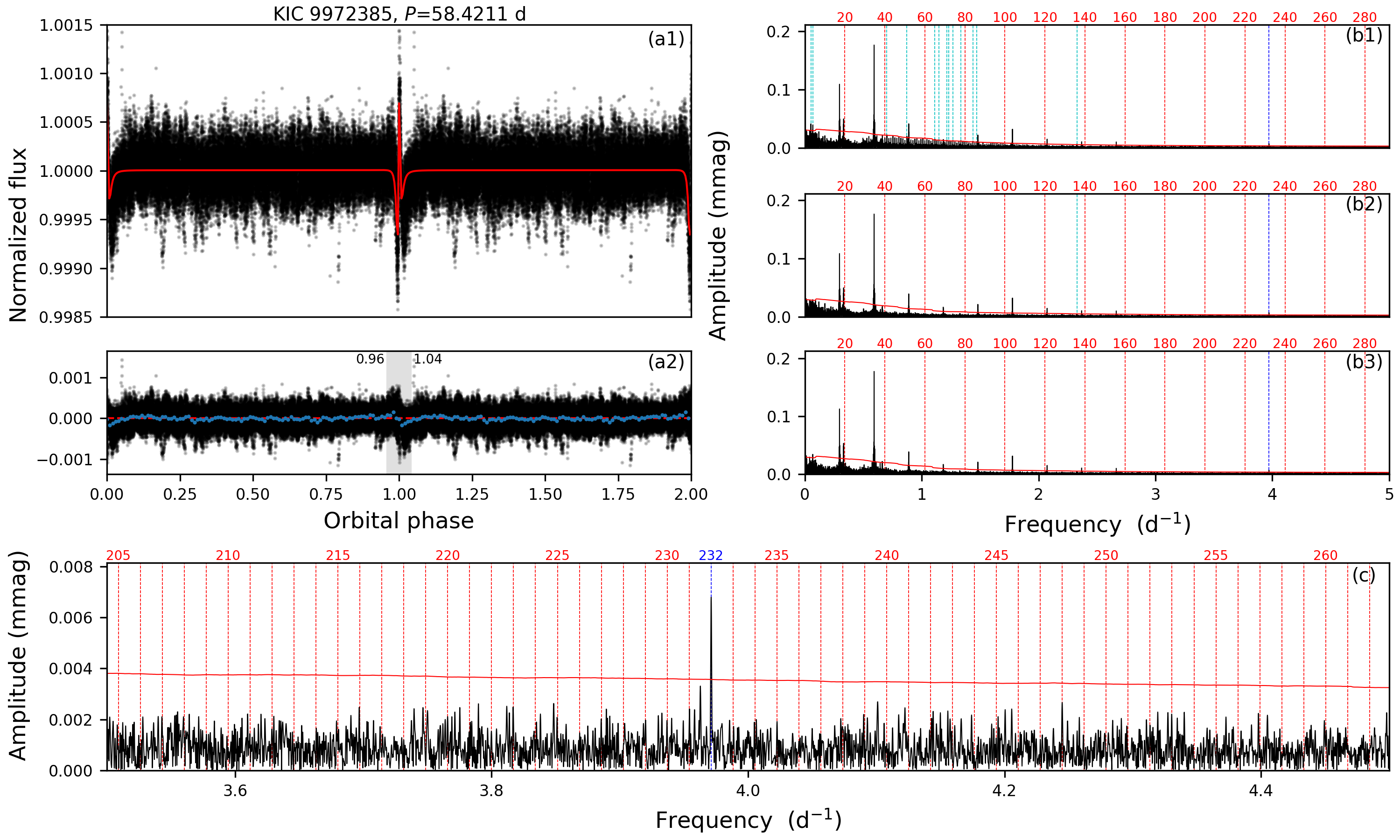}
	\caption{The analytic procedure for KIC 9972385. Panel (c) zooms in on panel (b3). The TEO candidate is the $n$ = 232 harmonic.
		\label{fig:9972385-ft}}
\end{figure}

\begin{figure}
	\centering
	\includegraphics[width=0.75\textwidth]{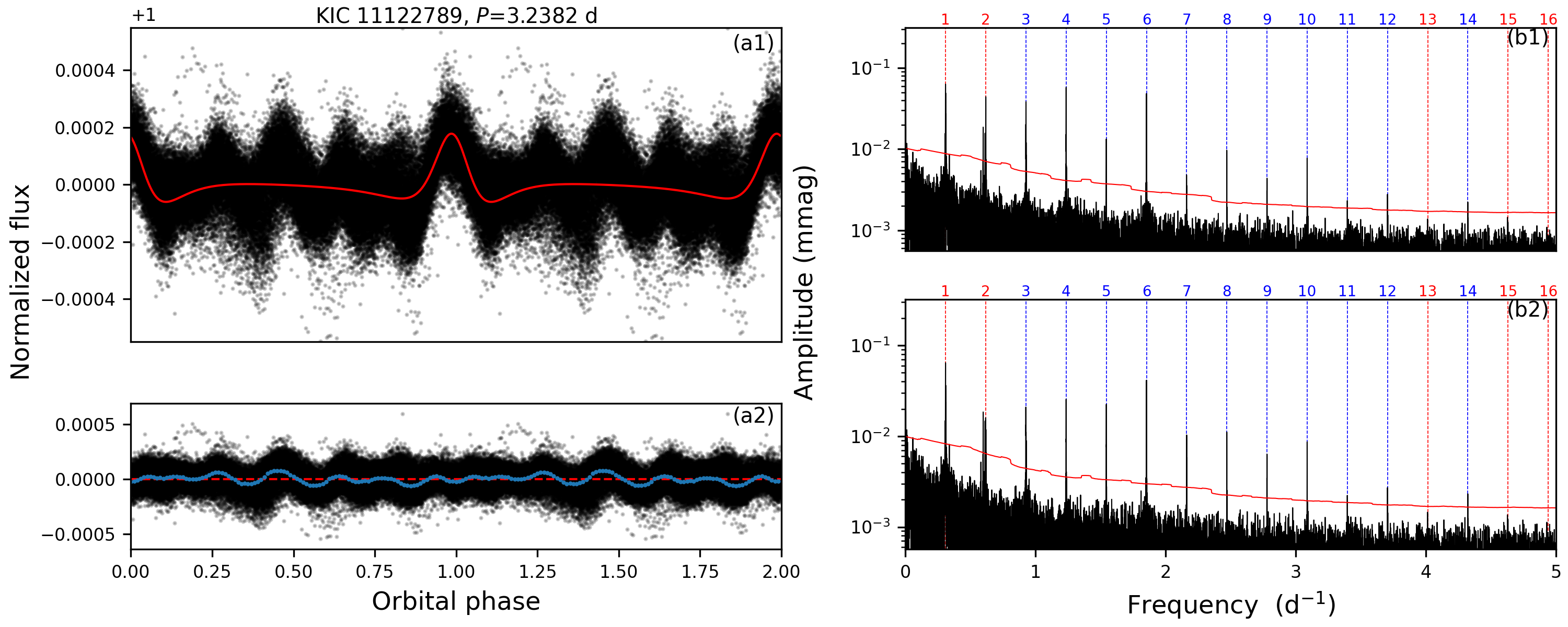}
	\caption{The analytic procedure for KIC 11122789. The TEO candidates are the $n$ = 6, 4, 5, 3, 8, 7, 10, 9, 12, 14, and 11 harmonics.
		\label{fig:11122789-ft}}
\end{figure}
\begin{figure}
	\centering
	\includegraphics[width=0.75\textwidth]{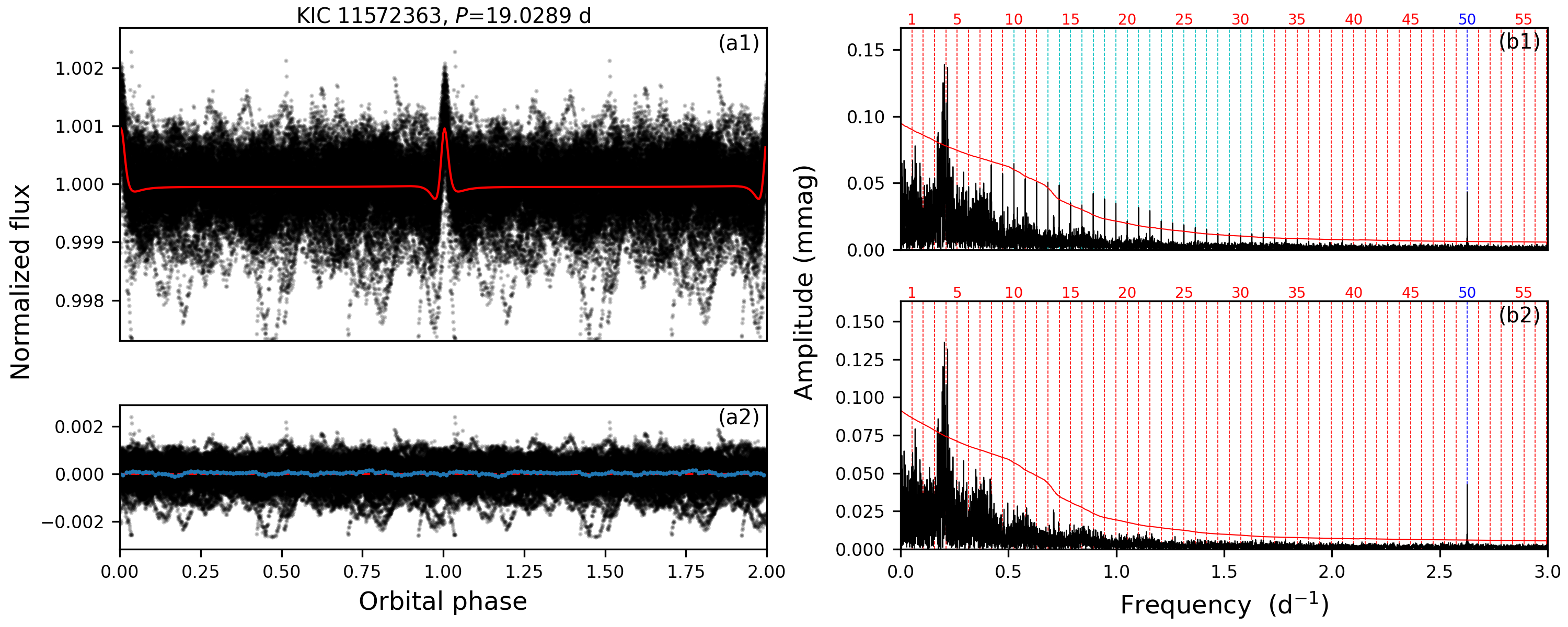}
	\caption{The analytic procedure for KIC 11572363. The TEO candidate is the $n$ = 50 harmonic.
		\label{fig:11572363-ft}}
\end{figure}



\end{document}